\documentclass[conference,compsoc]{IEEEtran}
\usepackage{cite}
\usepackage{amsmath,amssymb,amsfonts}
\usepackage{graphicx}
\usepackage{textcomp}
\usepackage{xcolor}
\usepackage{booktabs}
\usepackage{array}
\usepackage{makecell}
\usepackage{orcidlink}
\hypersetup{pdftitle={SweepSketch: Detecting Carpet-Bombing Attacks Before They End},
            pdfauthor={Yutong Yan, Sijia Du, Haowei Wang, Bo Yang}}
\newcolumntype{P}[1]{>{\raggedright\arraybackslash}p{#1}}
\def\BibTeX{{\rm B\kern-.05em{\sc i\kern-.025em b}\kern-.08em
    T\kern-.1667em\lower.7ex\hbox{E}\kern-.125emX}}
\begin{document}

\title{SweepSketch: Detecting Carpet-Bombing Attacks Before They End}

\author{\IEEEauthorblockN{Yutong Yan\textsuperscript{1,2}\orcidlink{0009-0003-6145-3898}}
\IEEEauthorblockA{\textsuperscript{1}\textit{University of Jinan} \\
\textsuperscript{2}\textit{Quan Cheng Laboratory} \\
Jinan, China \\
yanyt5@stu.ujn.edu.cn}
\and
\IEEEauthorblockN{Sijia Du\textsuperscript{1}\orcidlink{0009-0008-7437-3578}}
\IEEEauthorblockA{\textit{University of Jinan} \\
Jinan, China \\
dusj@stu.ujn.edu.cn}
\and
\IEEEauthorblockN{Haowei Wang\textsuperscript{2}\orcidlink{0000-0003-2481-4159}}
\IEEEauthorblockA{\textit{Quan Cheng Laboratory} \\
Jinan, China \\
sr-wanghw@qcl.edu.cn}
\and
\IEEEauthorblockN{Bo Yang\textsuperscript{2,1,*}\orcidlink{0000-0002-6866-6082}}
\IEEEauthorblockA{\textit{Quan Cheng Laboratory} \\
\textit{University of Jinan} \\
yangbo@ujn.edu.cn}
}

\maketitle

\begin{abstract}
Carpet-bombing DDoS attacks spread traffic almost uniformly across a destination prefix, keeping each host below alarm thresholds while exhausting prefix-level resources; spoofed sources make source-side evidence unreliable. Existing detectors either decide on windows of seconds, alarming only at window boundaries, or depend on supervised training, so short attacks often end before the first alarm. We present SweepSketch, a destination-anchored, per-packet sketch detector that keeps no source-side state, needs no training, and alarms while the attack is still in progress. Its core is a tagged self-cleaning windowed HyperLogLog (T-HLL) that turns uniformity testing into one event per distinct destination per window; per-packet cumulative-sum (CUSUM) decisions and dual exponentially weighted moving average (EWMA) gates form the decision chain. The reference state occupies 512 KB, and a bit-exact C++ implementation reaches 13.2 Mpps per core. On attacks lasting only 0.5 s, it raises its first alarm before the attack ends in 95.6\% of attacks; on real captures, 49.3\% alarm within the first second; static and mixed-traffic F1 both reach 0.991. SweepSketch is deployed in two hardware forms: a Tofino2 program carries the complete per-packet state chain and places the final gates on the control plane, while a DPU micro-engine keeps the entire decision chain in the fast path.
\end{abstract}


\section{Introduction}
\label{sec:intro}

Carpet bombing (sweep attack) is a distributed denial-of-service (DDoS) attack that spreads traffic nearly uniformly across a destination IP prefix, or several adjacent prefixes: every host receives only a few packets, far below single-host alarm thresholds, while prefix-level resources are exhausted wholesale. Attackers typically launch such attacks with spoofed source addresses or through reflection and amplification.

An industry analysis from 2023 \cite{b1} reports that a single sweep peaks at 200--600 Class C prefixes (/24s); 73.19\% of attacks are low-rate sweeps that consume bandwidth through the sheer number of prefixes, and 43.26\% of Class C-prefix attacks last under 5 minutes. Carrier-grade flow detection responds at minute scale: attackers deliberately keep per-destination traffic below thresholds and spread it uniformly, defeating per-IP rate detectors and per-source connection detectors alike.

This attack form poses three difficulties for detection. First, source-side evidence is unavailable. Existing spread detectors \cite{b2,b3,b4,b5} report per-host statistics: TRW \cite{b2} depends on per-source connection behavior, and the cardinality sketches \cite{b3,b4,b5} count the distinct peers of a single host, whether that host is a source or a destination. The signal of a uniform sweep lies in the distribution over the whole prefix, and spoofing removes the per-source state these methods rely on. They degrade sharply in this setting (Sections~\ref{sec:absence} and~\ref{sec:static}). Second, static distributions cannot separate attack from benign. Uniformly spread benign prefixes such as content delivery networks (CDNs) and anycast look identical in static distribution. ``Destinations in the prefix are dispersed'' therefore decides nothing; the temporal change signal is the main usable discriminator. Third, detection latency is decisive. NetRadar \cite{b6} triggers detection on a per-second cycle and identifies victim hosts from the previous second's features. An attacker can evade by switching targets within one detection cycle. GMCB \cite{b7} targets this sub-five-minute regime, where minute-scale response is too slow. DoLLM \cite{b8} reports that large language model inference latency limits real-time effectiveness. None of these detectors satisfies all three.

We propose SweepSketch, a sketch-based detector for carpet bombing, built on one principle: \textbf{testing, not estimation}. Existing sketch detectors target superspreaders or traffic volume and do not address the decision requirements of a uniform sweep; SweepSketch anchors detection at the destination and decides per packet. Entropy estimators must store counters whose width scales with $\log n$ per bucket, whereas uniformity testing needs only decision accumulators whose widths scale with $\log(1/\varepsilon)$ and $\log(1/\alpha)$. We realize this in 44-byte fixed-width buckets. The core primitive, T-HLL, is a register-packed windowed-HLL variant. Its signature mechanism is \textbf{tagged self-cleaning}: window clearing is encoded into the tag itself; stale data expires with no clearing operation. Per-packet CUSUM (cumulative sum) decisions and dual-EWMA (exponentially weighted moving average) change gates complete the decision chain, and a two-level structure (/24 and /16, denoted L1 and L2) covers single-prefix sweeps and staggered /22--/20 attacks (traffic split across adjacent prefixes). The per-packet cost is 3 hashes and 4--6 register reads/writes per bucket, with no division, no floating point, and no per-packet timestamps.

The contributions are fourfold:
\begin{enumerate}
\item \textbf{T-HLL: a tagged self-cleaning windowed HLL} (Section~\ref{sec:thll}). It encodes window clearing into the 3-bit tag, so stale data expires without clearing; one pass over 32 8-bit registers yields the collision statistic and the per-window distinct count, turning uniformity testing into a per-window event stream in 44-byte buckets. To our knowledge, no prior windowed HLL variant encodes window clearing into the tag to eliminate explicit clearing.
\item \textbf{SweepSketch: the first per-packet sketch detector for carpet bombing} (Sections~\ref{sec:design}--\ref{sec:theory}). It combines per-packet CUSUM, dual EWMA gates, and a /24--/16 hierarchy with bucket aging, confirmation, and cross-prefix correlation; it needs no training and keeps its state in 512 KB.
\item \textbf{Damage-bounded detection and evaluation} (Sections~\ref{sec:lowrate}, \ref{sec:deadline}, and~\ref{sec:eval}). We define damage at alarm and deadline-conditional recall $R(\delta)$, and prove a latency--damage floor: waiting for $k$ new destinations gives latency $\ge k/\lambda$ and damage $\ge k/S$, while window-batch detectors reach damage 1.0 for sub-window attacks. SweepSketch's 12-destination gate is close to this floor; on attacks lasting only 0.5 s it alarms before the attack ends in 95.6\% of attacks. On 71 real captures and 30 holdout files, static and mixed-traffic F1 is 0.991 (episode-level alarm precision 0.89); 49.3\% of real captures alarm within 1 s, and cross-prefix correlation catches a burst-and-rotate attacker at window 2.
\item \textbf{Two hardware deployment forms} (Section~\ref{sec:deploy}). We deploy the complete per-packet state chain on a Tofino2 switch (90 tables, 12 stages, 500 KB) as a two-pass program, with the final gates and confirmation on the control plane. The complete decision chain is deployed in the fast path of a DPU micro-engine (1,135 instructions, 240 KB). Both forms implement the same decision function, and the micro-engine form is bit-identical to the reference on 10.47 M packets.
\end{enumerate}

\begin{figure*}[t]
\centering
\includegraphics[width=0.95\textwidth]{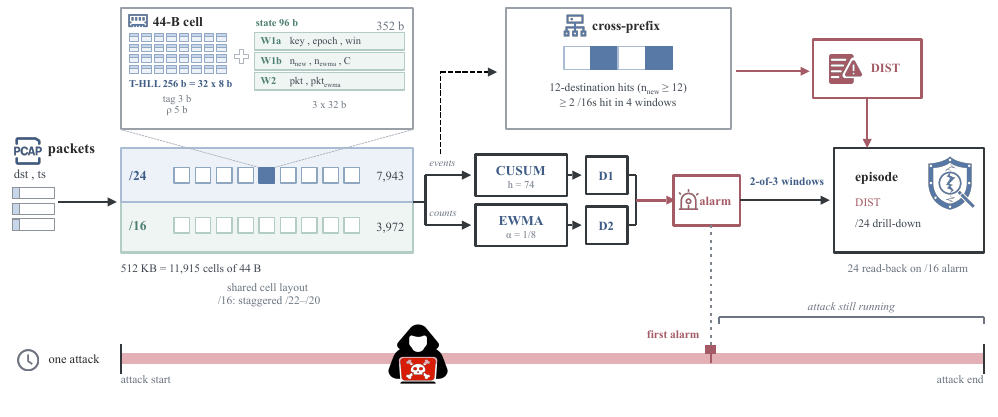}
\caption{SweepSketch architecture: every packet keys /24 and /16 cells (44 B each, 512 KB in total); T-HLL events feed the per-packet CUSUM and EWMA gates, and confirmed alarms carry a cross-prefix DIST alarm and a /24 drill-down. The timeline (bottom) shows the first alarm before the attack ends.}
\label{fig:arch}
\end{figure*}

\section{Background and Threat Model}
\label{sec:background}

\subsection{Carpet-Bombing Attacks}
\label{sec:attacks}

The attacker picks one destination IP prefix (usually a /24) or several adjacent ones, and spreads traffic nearly uniformly over all addresses in the prefixes, so each destination receives only a few packets. In staggered attacks, traffic is distributed across adjacent /22--/20 prefixes, each at lower intensity. The target is prefix-level resources (connection-table entries, aggregate bandwidth, mitigation systems, operator attention), not any single host. The earliest formal academic characterization is the PAM'21 multiprotocol carpet-bombing study \cite{b9}. The industry definition (NETSCOUT/Arbor) largely agrees: the attack fills every IP of a /24, each IP's rate is deliberately kept below the mitigation threshold, and source addresses are spoofed; the industry report \cite{b1} describes it as ``attacking multiple IP addresses within one or more Class C IP address segments simultaneously''. Honeypot observations \cite{b10} separate its signatures from single-target reflection DDoS.

The 71 real attack files fall into three forms (Section~\ref{sec:setup}):
\begin{itemize}
\item \textbf{Scan-type} (SYN/ACK/WEB/UDP/HTTP, median 1--7 packets/flow): strong flow-level, weak packet-level evidence.
\item \textbf{Flood-type} (ICMP, median 367 packets/flow): grouped bursts, strong packet-level evidence.
\item \textbf{Reflection-type} (DNS/HTTP/HTTPS/TCP, median 1 packet/flow): reflection further erodes source-side evidence.
\end{itemize}

All three forms share the same destination-side signature (near-uniform spreading over the prefix's destinations), which makes destination-side uniformity the unified signal. Uniform benign prefixes such as CDNs and anycast are the family a detector must suppress.

\subsection{Threat Model}
\label{sec:threat}

We adopt the following threat model:
\begin{itemize}
\item \textbf{Attacker capabilities}: (a) spoof source addresses arbitrarily, or launch attacks through reflection amplifiers; (b) choose the destination prefix and granularity (a single /24 or staggered /22--/20), tuning the rate and packets per flow per prefix; (c) choose the packet ordering (per-packet rotation or per-destination bursts); (d) know the detector's complete design and all parameters (Kerckhoffs's assumption) and construct orderings specifically to evade detection.
\item \textbf{Detector scope}: the detector observes only destination addresses and timestamps and keeps no per-source state. Parameters are calibrated from benign traffic only, and attack data are used only for evaluation (Section~\ref{sec:params}); the detection floors (Section~\ref{sec:theory}) delineate the undetectable region.
\item \textbf{Deployment location}: edge switches at campus, data-center, or enterprise borders, covering 5k--20k active /24s (scale limits in Section~\ref{sec:scale}).
\end{itemize}

\subsection{The Structural Absence of Source-Side Evidence}
\label{sec:absence}

\begin{figure}[t]
\centering
\includegraphics[width=\columnwidth]{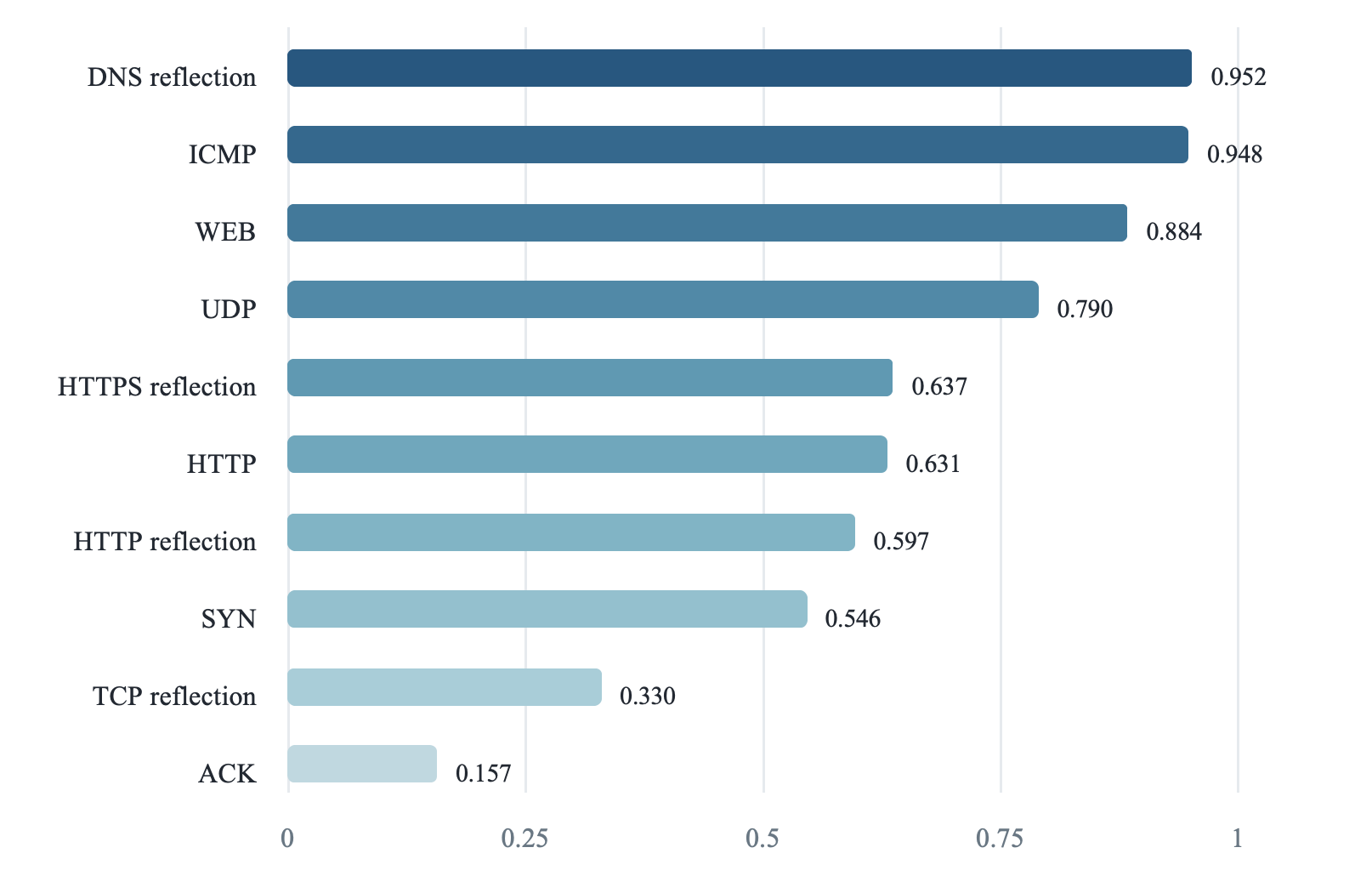}
\caption{Source aggregation ratio of the 10 attack types.}
\label{fig:sourceagg}
\end{figure}

For the 10 attack types, the source aggregation ratio $\nu = |Q_{/24}|/|Q|$ (distinct source /24s over distinct source IPs) ranges from 0.157 (ACK) to 0.952 (DNS reflection); Figure~\ref{fig:sourceagg} shows the per-type values. Reflection sources span nearly the whole address space, and scan sources are also widely dispersed. Source-side instantiations of the baseline methods show the consequence: they reach at most F1 0.049 over the whole corpus (Section~\ref{sec:static}), fall to near zero after per-packet source rewriting, and recall only 0.21\%--5.63\% on the holdout (Section~\ref{sec:real}). Our detector therefore anchors at the destination side. Destination-side methods use /24 prefixes as ground truth (GT); source-side methods use all source IPs in the file, and the two conventions are labeled in each table.

\subsection{Fixed-Width Memory and the Testing-Not-Estimation Principle}
\label{sec:testing}

Under the fixed-width constraints of the threat model, estimation and testing are not equally expensive. Single-pass streaming additive entropy estimation has a memory lower bound of $\Omega(\varepsilon^{-2}/\log(\varepsilon^{-1}))$ bits \cite{b11}, and the bounded-memory entropy sketch \cite{b12} has counters that scale with $\log n$; its memory grows with the precision requirement. Uniformity testing needs only a collision statistic (one event per distinct destination per window) plus a decision accumulator whose width scales with $\log(1/\alpha)$. Our 352-bit cell's $\varepsilon$-dependent field is 5 bits, against an estimation-side bound of hundreds of bits at $\varepsilon = 0.05$; the design exploits this asymmetry rather than a matched lower bound.

\section{Design Overview}
\label{sec:design}

SweepSketch is organized around three principles:
\begin{itemize}
\item \textbf{P1 Testing, not estimation}: buckets hold only decision accumulators (widths scaling with log thresholds), never counters of $\log n$ width.
\item \textbf{P2 Decisions at the bucket, per packet}: each packet's decision is a function of that packet and the state of its own bucket, with no global counters, no cross-bucket state, and no decision deferred to a window boundary. Everything downstream of the bucket consumes state records or alarm records rather than packets.
\item \textbf{P3 Time is per bucket}: window rollover and the EWMA baseline are carried per bucket and advanced lazily when the bucket is touched, so a change gate compares the current window against that bucket's own baseline. Neither a global sweep nor an external time layer is needed.
\end{itemize}

These principles constrain the decision, independently of where it executes. The reference state machine keeps all 44 B in the bucket and evaluates every comparison on each packet. A deployment may keep only the per-packet accumulators on the fast path and place the per-window baselines and the final gate conjunctions on a control plane that receives one state record per packet, or it may keep the entire decision chain in a general-purpose fast path. Section~\ref{sec:plan} describes the three forms and evaluates the cost of the split; the decision function is the same in all of them.

Batch decisions at an epoch boundary impose at least one window of latency, and entropy estimation forfeits the $\varepsilon^{-2}$ advantage (Section~\ref{sec:testing}). Sequential testing makes each packet a decision rather than a sample, which is what a sub-second deadline requires; latency drops to packet level and the Wald--Wolfowitz theory \cite{b13} applies. SweepSketch adopts this form: the epoch ($2^{32}$ ns = 4.29 s, the wrap of the low 32 bits of the ingress timestamp; Appendix~\ref{app:notation}) sets only counting and baseline-reset granularity, and decisions remain per packet. Four components address the three difficulties (Section~\ref{sec:intro}) and the threat model's granularity freedom (Section~\ref{sec:threat}): T-HLL (statistic), CUSUM (decisions), dual-EWMA gates (change semantics), and the /24--/16 hierarchy (granularity), Sections~\ref{sec:thll}--\ref{sec:hierarchy}.

Figure~\ref{fig:arch} shows the 44-byte bucket and the four components that keep state in it. Per packet, the destination address is parsed into two bucket keys (the /24 and /16 prefixes). A seven-stage pipeline then runs: S1 hashing (bucket index $\times$2 + T-HLL $r$/$\rho$ $\times$1) $\to$ S2 T-HLL event decision (tag mismatch $\lor$ larger $\rho$) $\to$ S3 lazy window rollover $\to$ S4 LLR increment table lookup (2 entries) $\to$ S5 state update $\to$ S6 raw-alarm predicate (D1 dispersion-channel gate + D2 volume-channel gate; evaluated where each form places it, Section~\ref{sec:plan}) $\to$ S7 snapshot reporting ($\le$48 B/packet in the design; 28 B in the compiled deployment). Alarm digests are deduplicated and aggregated by prefix into operator alerts, with on-demand forensics to locate the swept prefixes.

\section{Detailed Design}
\label{sec:detailed}

\subsection{Bucket Layout and Memory Accounting}
\label{sec:bucket}

Each bucket (a fixed-width cell in the theory) is 44 B, and both levels share the reference layout: a 32$\times$8-bit T-HLL array plus three 32-bit words W1a (key, epoch, win), W1b ($n_{\mathrm{new}}$, $n_{\mathrm{ewma}}$, $C$), and W2 ($pkt$, $pkt_{\mathrm{ewma}}$). The 512 KB reference account yields 11,915 bucket slots (7,943 at the /24 level and 3,972 at /16; 11,915$\times$44 B = 524,260 B, leaving 28 B), 5.8$\times$ the /24 slot count of the entropy baseline at the same 512 KB (2,048 slots of $z=64$ 32-bit counters; Appendix~\ref{app:entropy}). The compiled register file rounds the two depths to 8,192 and 4,096 slots and packs 42 B per /24 entry and 41 B per /16 entry (500 KB) with the deployed field split; the compiled form keeps the per-packet subset of these fields, and Appendix~\ref{app:deploy-meth} states which decisions use the remaining fields. Collisions follow the cold/stale replacement rule of Section~\ref{sec:lifecycle}. 

\subsection{T-HLL: A Tagged Self-Cleaning Collision Statistic}
\label{sec:thll}

Implementing testing-not-estimation (P1) faces one difficulty: the statistic that uniformity testing needs is an event stream in which each distinct destination fires exactly one event per window (first touch counts, repeat touches do not), maintained exactly, per window, per packet.

Neither off-the-shelf primitive suffices: HLL is a global estimator with relative error and no windowing. Its noise would feed the CUSUM decision; a 256-bit bitmap per /24 records exactly, but lazy clearing (255 bits on a new window's first touch) costs 32 reads/writes on one packet, beyond the 4--6 RMW budget; eager clearing (an external per-window sweep) moves temporal semantics outside the bucket, violating P3. The register array offers only unconditional single-value overwrite, with no global clear instruction. T-HLL therefore maintains the stream exactly under a 32$\times$8-bit fixed-width array and a 4--6 RMW budget.

T-HLL encodes clearing into the tag itself; we call this mechanism \textbf{tagged self-cleaning}. Each register is 8 bits = \{win tag 3b $|$ $\rho$ 5b\}, with single-value overwrite as the only write rule: on a tag mismatch (this window's first touch of the register) it writes T-HLL$[r] \leftarrow \{win,\ \rho\}$, resetting the baseline, and on a hit it writes $\{win,\ \rho\}$ only if $\rho > \rho_{\mathrm{old}}$. The event decision is $\mathrm{event} = (tag \ne win) \lor (\rho > \rho_{\mathrm{old}})$. Because the tag refreshes on every write, the first touch of a register in a window always fires and later touches fire only on a strictly larger $\rho$; when the next window refreshes the tag, stale tags mismatch again. Each distinct destination thus contributes exactly one first-touch event per window, and $n_{\mathrm{new}}$ is the exact event count: under the injection mapping (distinct destinations map injectively to the 32 registers, each carrying at most one), it equals the exact distinct count. Collision loss under the production hash is in Lemma~L1 (Section~\ref{sec:lemmas}). At bucket creation the registers start at \{tag 0, $\rho$ 0\} with $win = 1$, so the first window fires all events with no cold-start blind window; on prefix replacement win inherits the old value plus 1 (Appendix~\ref{app:convb}), which likewise guarantees tag mismatch.

The event model thereby reduces to an idealized Bernoulli process: one first-touch event per distinct destination, plus possible strict-$\rho$ record events among repeated hits (a geometric tie-breaking record process with tie probability 1/3). 

Two measured effects limit the ``exact'' reading of $n_{\mathrm{new}}$. First, under the production hash the event count is biased by register collisions: relative to the exact distinct count, an exact-count comparison measures $-7.3\%$ at $S=4$, $-14.2\%$ at $S=16$, and $-25.2\%$ at $S=32$; only the injection mapping is exact. Second, the 3-bit tag wraps every eight active windows. A host that returns exactly every eight windows can be re-counted as fresh; in a workload in which 25\% of hosts return every eight windows, 2.6--40.9\% of return events are missed (20.4\% at $S=8$).

\subsection{The CUSUM Statistic and LLR Increment Table}
\label{sec:cusum}

With the statistic supplied by T-HLL, the dispersion channel D1 still faces the decision problem: making fast, provable decisions from the event stream under a per-packet constant budget, with no floating point and no division. On an idealized Bernoulli event stream, SPRT (the sequential probability ratio test) is optimal in expected sample size for simple hypothesis pairs (Wald--Wolfowitz \cite{b13}), but its log-likelihood ratio is real-valued and cannot be expressed directly. CUSUM is its fixed-point-expressible reflected-random-walk form \cite{b14}: $C\leftarrow\min(C_{\max},\max(0,C+\mathrm{LLR}[\mathrm{event}]))$ with $C_{\max}=2^{16}-1$, and the LLR table has two compile-time entries on an 8-bit fixed-point scale, $z_+=\mathrm{round}(\log_2(\theta_1/\theta_0)\cdot8)=38$ and $z_-=\mathrm{round}(\log_2((1-\theta_1)/(1-\theta_0))\cdot8)=-26$.
The rates are $\theta_0 = 0.0345$ (Section~\ref{sec:params}) for benign traffic and $\theta_1 = 0.9$ for attack traffic (tag mismatch always fires, minus a collision-suppression margin, Section~\ref{sec:lemmas}). A benign bucket drifts at $\mu_0 \approx -23.8$ per packet, so $C$ stays near 0; an attack bucket drifts at $\mu_1 \approx +31.6$ and, under a per-packet rotation ordering, crosses $h = 74$ on the second packet. The event fraction must therefore exceed $|z_-|/(z_+ + |z_-|) = 26/64 \approx 41\%$ to sustain positive drift, which a per-packet rotation sweep (event fraction $\approx 1$) exceeds and a grouped-burst ordering does not (the ordering floor of T6, Section~\ref{sec:t4t6}).

\subsection{Dual-EWMA Gates and Alarm Semantics}
\label{sec:gates}

A CUSUM threshold crossing shows that a deviation happened, not that it is new. Benign prefixes that stay uniform over time (Section~\ref{sec:attacks}) cross it too, so the discriminating signal must be temporal. The dual-EWMA gates express that change semantics inside fixed-width buckets, with lazy per-window rollover and no external time layer. An alarm is raised by either the D1 dispersion channel (a CUSUM crossing plus the D1 gate) or the D2 volume channel:
\begin{align*}
\mathrm{alarm1} ={} & \bigl[C \ge 74\bigr] \land \bigl[n_{\mathrm{new}} \ge \\
& \max\bigl(8,\ 2\,n_{\mathrm{ewma}},\ 12\cdot\mathbb{1}[n_{\mathrm{ewma}}<4]\bigr)\bigr]
\end{align*}
\begin{align*}
\mathrm{alarm2} ={} & \bigl[pkt \ge 200\bigr] \land \bigl[pkt \ge 3\,pkt_{\mathrm{ewma}}\bigr] \land \\
& \bigl[n_{\mathrm{new}} \ge \max\bigl(6,\ 2\,n_{\mathrm{ewma}}\bigr)\bigr] \land d2_{\mathrm{held}}
\end{align*}

Write $G = \max(S_{\mathrm{min}}, 2\,n_{\mathrm{ewma}}, S_{\mathrm{quiet}})$ for the D1 gate. The dispersion floor $S_{\mathrm{min}} = 8$ is the minimum of $G$ and is attained only at $n_{\mathrm{ewma}} = 4$; under the injection mapping this allows an alarm as early as the 8th new-destination packet. $G$ steps down once and then rises. Under the production hash, the single-window detection rate is T5's $\delta(S)$. The change gate separates a long-term-uniform CDN from an attack that turns uniform. The quiet floor $S_{\mathrm{quiet}} = 12\cdot\mathbb{1}[n_{\mathrm{ewma}} < 4]$ closes the cold-start gap, where a new prefix has a baseline near zero and the $2\,n_{\mathrm{ewma}}$ threshold would not bind. The D2 volume detector covers flood-type attacks through the packet channel: $P_{\mathrm{MIN}} = 200$ is the rate floor, the 3$\times$ EWMA gate expresses volume change, $K_{\mathrm{FLOOD}} = 6$ is the spread lower bound, and $d2_{\mathrm{held}}$ requires the gate to have held over the preceding window, filtering one-off benign fluctuations. Concretely, $d2_{\mathrm{held}}$ is a one-bit latch set at window rollover from the closed window's final values, as $d2_{\mathrm{held}} = [\,pkt \ge 200 \land pkt \ge 3\,pkt_{\mathrm{ewma}} \land n_{\mathrm{new}} \ge \max(6,\, 2\,n_{\mathrm{ewma}})\,]$; since $pkt$ and $n_{\mathrm{new}}$ only grow within a window, this equals the gate holding at any point in the closed window. The EWMA updates at rollover as $x \leftarrow x + (n_{\mathrm{new}}-x) \gg 3$ (absorption rate 1/8), so the baseline rises as it absorbs the attack, converging to the attack level after roughly 5.2--7 windows, at which point the gate closes by itself (proposition T4, Section~\ref{sec:t4t6}); after the attack ends the baseline recedes and re-detection capability returns. Window rollover is lazy per bucket, so no external sweep is needed. Both predicates are functions of the bucket's own state; Section~\ref{sec:plan} states where each form evaluates them.

\subsection{Bucket Lifecycle, Confirmation, and Cross-Prefix Correlation}
\label{sec:lifecycle}

The 44-byte bucket has no spare field for aging: its 5-bit epoch is already the last-active epoch. A colliding key replaces the incumbent if the incumbent is cold (low packet count and no CUSUM evidence, i.e., $pkt<P_{\mathrm{MIN}}$ and $C<h$) or stale (not touched for $\mathrm{AGE\_WIN}$ windows). The thresholds $P_{\mathrm{MIN}}$, $h=74$, and $\mathrm{AGE\_WIN}=4$ are in Table~\ref{tab:params} and Section~\ref{sec:cusum} (4 windows $\approx$17.2 s). Because resetting a bucket restarts its CUSUM, aging raises raw alarm churn. The control plane therefore runs confirmation: D2 must hold in two of the last three windows; repeated alarms on one prefix are cooled for 20 windows; and a prefix that alarms three times in two hours without sustained $n_{\mathrm{new}}$ growth is quarantined for 24 hours. The operator-facing stream is \emph{episodes}, not raw alarms.

Cross-prefix correlation records, for each /16, the last $W=4$ windows in which any child /24 first crossed $n_{\mathrm{new}}\ge12$ (the 12-destination signal, independent of which level's gate fired). At least $K=2$ distinct /16s within the last $W=4$ windows raise a DIST (cross-prefix) alarm. It stores $W$ bits per /16, i.e., linear in the number of /16 prefixes on the control plane; the data-plane bucket remains 44 bytes. Confirmation, cooling, quarantine, and cross-prefix correlation are control-plane functions over the same digest stream.

\subsection{Two-Level Hierarchy}
\label{sec:hierarchy}

Because the attacker chooses the prefix and granularity (Section~\ref{sec:threat}), uniform spreading over a single /24 clears the D1 gate $G$, but staggered /22--/20 attacks spread traffic over adjacent /24s, hitting only a few destinations per /24, each below the gate on its own, which leaves any single-level monitor with a structural blind spot that an attacker can construct. The two-level hierarchy answers this structurally, covering the attacker's whole granularity spectrum under one fixed memory budget. The /24 level targets uniform spreading within a prefix; the /16 level provides an aggregated view with the same T-HLL structure and the same test at a different resolution, so an attack with only a few hits per /24 aggregates at /16 to a distinct total above the floor. The cost of aggregation is quantifiable: /16 buckets carry more distinct destinations, and register-collision suppression lifts the effective gate from 8 to 9 (quiet floor 12 to 14) through the second-order term of Lemma L1 (Section~\ref{sec:lemmas}). On a /16 alarm, the active /24 buckets under that /16 are read to locate the swept prefixes (forensics, not detection); the decision signal itself is bucket-local, and Section~\ref{sec:plan} states where it is evaluated.

\subsection{Parameters and Calibration}
\label{sec:params}

Table~\ref{tab:params} lists the frozen parameters and their provenance.

\begin{table*}[t]
\centering
\caption{Parameters, values, and provenance.}
\label{tab:params}
\begin{tabular}{@{}P{3.3cm}P{2.1cm}P{11.2cm}@{}}
\toprule
Parameter & Value & Provenance \\
\midrule
$\theta_0$ & 0.0345 & Upper median of per-file medians over the 10-capture benign calibration corpus, with 1-in-10 packet sampling (frozen before evaluation); the median depends on the backbone trace used (Appendix~\ref{app:data}) \\
$\theta_1$ & 0.9 & Tag mismatch always fires $-$ collision-suppression margin (Lemma L2: event rate $\ge 0.927$ at $S=8$) \\
LLR table & \{+38, $-26$\} & Precomputed from $\theta_0$/$\theta_1$ at compile time \\
$h$ (CUSUM threshold) & 74 & Design freeze; T2 evaluates \\
$S_{\mathrm{min}}$ / $S_{\mathrm{quiet}}$ (dispersion/quiet floors) & 8 / 12 & Gate min. / cold-start \\
$\mathrm{EWMA\_SHIFT}$ / $\mathrm{D2\_MULT}$ (absorption rate / volume-gate multiple) & 3 / 3 & T4 absorption / volume margin \\
$K_{\mathrm{FLOOD}}$ (spread lower bound) & 6 & Lemma T6 over-concentration floor \\
$P_{\mathrm{MIN}}$ & 200 & Lemma T6 rate floor (implemented $\lambda_{\min}=200/4.29\approx47$ pps; the 40 pps design target is below it; Appendix~\ref{app:data}) \\
$\mathrm{AGE\_WIN}$ & 4 & Bucket is replaceable if $((g-b.\mathrm{epoch})\bmod32)\ge4$ (no extra bits; epoch is already stored) \\
$\mathrm{COOLDOWN\_WIN}$ & 20 & Episode cooling ($\approx86$ s); D2 needs 2 of the last 3 windows \\
$Q_{\max K}$ / $Q_{\mathrm{growth}}$ & 16 / 1.25 & Quarantine suppressed if any episode's $n_{\mathrm{new}}>16$ (benign max 15) or $n_{\mathrm{new}}$ growth $\ge1.25\times$ over the last three episodes \\
$W$ / $K$ (cross-prefix) & 4 / 2 & Distinct /16s with a 12-destination hit in the last 4 windows \\
44 B/cell (reference) & 11,915 cells, 512 KB (deployed 8,192 + 4,096) & bit-width accounting; $N_{L1}{:}N_{L2} = 2{:}1$ (7,943 /24 + 3,972 /16) \\
\bottomrule
\end{tabular}
\end{table*}

\textbf{Calibration discipline}: $\theta_0$ is taken only from the benign side; attack data serve only for validation. The calibration set \{IoT-Benign1, IoT-Benign2\} is separated from the test set \{IoT-Benign, IoT-Benign3\}, parameters were frozen on a development set (CIC plus the strategy grid), and the 30 holdout files were never used for tuning. Appendix~\ref{app:data} reports the $\theta_0$ provenance and sensitivity; Appendix~\ref{app:proofs} maps each parameter to its proposition and Monte Carlo check.

\section{Theoretical Analysis}
\label{sec:theory}

\subsection{Event Model and Assumptions}
\label{sec:eventmodel}

\textbf{Setting and assumptions.} Destinations are drawn independently from $D$ within a window, and $H(\mathrm{dst})$ is a uniform 32-bit Bob Jenkins hash; the register index is $r=H(\mathrm{dst})\ \&\ 31$ and $\rho$ is the leading-zero count of $H(\mathrm{dst})\gg5$. A mismatch writes $\{win,\rho\}$; after a hit the register holds $\max(\rho,\rho_{\mathrm{old}})$. (A1) Windows are i.i.d.: fresh-$\rho$ makes each window depend only on its packets and the previous tag. (A2) $\theta_0=0.0345$ is the upper median of per-file benign medians (Table~\ref{tab:params}). (A3) Under a uniform sweep every packet hits a new destination (median 1.0 packets/flow in the scan corpus), so the event rate is $\ge1-$ collision suppression $\ge\theta_1=0.9$.

\subsection{Basic Lemmas and Sample Complexity}
\label{sec:lemmas}

\textbf{Lemma L1 (T-HLL event counting and collision accounting)} $S$ distinct destinations in a window fall into $m = 32$ registers, and register $r$ is touched $k_r$ times. Ties in $\rho$ have positive probability ($\Pr(\rho_1=\rho_2)=1/3$). A hit counts as a record event only when $\rho$ strictly exceeds the record already in the register. Then:
\begin{enumerate}
\item Register $r$'s event count = 1 (first-touch tag mismatch) + the strict-$\rho$ records among hits, and
\[
\mathbb{E}[n_{\mathrm{new}} \mid S] \;\ge\; S - \frac{S(S-1)}{96},
\]
with the collision loss as the second-order term; the derivation is in Appendix~\ref{app:proofs}.
\item If $S \le 32$ and the mapping is injective, $n_{\mathrm{new}} = S$ pointwise.
\end{enumerate}

\textbf{Lemma L2 ($H_1$ event rate)} Under the injection mapping the event fraction is $1 \ge \theta_1$. Without it, when every packet hits a new destination, the event fraction at $S=8$ is $\ge 0.927 \ge \theta_1$ (Appendix~\ref{app:proofs}).

\textbf{Proposition T1 (sample complexity and statistical sufficiency)} Each bucket's $n_{\mathrm{new}}$ is a fixed-width realization of the destination distribution's collision statistic. For the ``uniform vs $\varepsilon$-far'' decision (total variation distance from uniform $\ge \varepsilon$) over the $m=32$ registers, the sample complexity is $5.7/\varepsilon^2$ samples \cite{b15,b16}. In a fixed-width cell, uniform spreading with $S \ge G$ meets the decision condition within $G$ packets, and $C$ crosses at the $\lceil h/z_+\rceil = 2$nd new-destination packet of a repetition-free ordering.

\subsection{Suppression and Detection Bounds}
\label{sec:bounds}

\textbf{Proposition T2 (false-positive suppression and detection latency)} For the reflected random walk $C_t=\max(0,C_{t-1}+Z_t)$ with $Z_t\in\{+38,-26\}$, (i) \emph{suppression}: $\Pr(\max_t C_t\ge h)\le e^{-\gamma^*h}$ with $\gamma^*=0.0857$ \cite{b17}, so at $h=74$ the per-window per-bucket bound is $1.8\times10^{-3}$; the dual-EWMA gates suppress further, and measured whole-system false positives appear in Section~\ref{sec:benign}. (ii) \emph{detection}: the $H_1$ drift is $\mu_1=\theta_1z_++(1-\theta_1)z_-=31.6$ per packet; Hoeffding \cite{b18} and Wald \cite{b19} give $\mathbb{E}[\tau]\le(h+z_+)/\mu_1=3.5$ packets, and with the $n_{\mathrm{new}}$ gate $\mathbb{E}[\mathrm{latency}]\le G+3.5$ packets; under the injection mapping the latency is exactly $G$ packets, because $C$ crosses at the second event and the gate passes once $G$ new destinations have been seen. (iii) \emph{fixed point and truncation}: fixed-point and truncation corrections are standard \cite{b20,b21,b22}; Appendix~\ref{app:proofs} gives the details.

\textbf{Proposition T3 (fixed-width realization).} The 44-byte cell implements the test with widths set by the thresholds rather than by the stream: 352 bits in total, all fields threshold-sized (Appendix~\ref{app:proofs}). No field scales with $n$, and the $\varepsilon$-dependent field is 5 bits. The false-alarm and latency guarantees are those of T2.

\subsection{Change Semantics, Coverage, and Detection Floors}
\label{sec:t4t6}

\textbf{Proposition T4.} For a stable benign prefix, $\mathbb{P}(n_{\mathrm{new}}\ge2n_{\mathrm{ewma}})\le(e/4)^{n_{\mathrm{ewma}}}\le e^{-n_{\mathrm{ewma}}/3}$; if the single-window false-positive rate is $p$, D2 confirmation reduces it to $O(p^2)$. The EWMA is a contraction with factor $7/8$, absorbs a step in $\le\log2/\log(8/7)\approx5.2$ windows (7 with integer shifts), so alarms stop at the new baseline and resume after the attack ends.

\textbf{Lemma T5.} If some /24 reaches $\max(S_{\min},2n_{\mathrm{ewma}},S_{\mathrm{quiet}})$, the /24 level alarms; otherwise the /16 aggregation may. Under the injection mapping a sweep with $S\ge G$ reaches $n_{\mathrm{new}}\ge G$ and $C\ge h$ within $G$ packets; under the production hash $\mathbb{E}[n_{\mathrm{new}}\mid S]\ge S-S(S-1)/96$, with single-window detection $\ge1-\delta(S)$ and multi-window $1-\delta(S)^W$ (Appendix~\ref{app:proofs}). Reporting granularity is off by at most one level.

\textbf{Lemma T6 (four floors).} (i) \emph{Rate}: D2 needs $pkt\ge200$ per window ($\approx47$ pps); D1 needs distinct $\ge\max(8,2n_{\mathrm{ewma}},12\cdot\mathbb{1}[n_{\mathrm{ewma}}<4])$. (ii) \emph{Ordering}: CUSUM needs an event fraction $>26/64\approx41\%$; bursts of $\ge3$ packets to the same destination cap $C$ at $38<74$, so D1 is silent and D2 must carry the attack while its rate and confirmation hold. (iii) \emph{Over-concentration}: floods with distinct/window $<K_{\mathrm{FLOOD}}=6$ trigger neither channel. (iv) \emph{Dense-window absorption}: staggered attack windows on the same target are absorbed by the EWMA baseline. Each floor follows from its gate; the constructions map to the strategy-grid cells and the injection-miss composition.

\section{Evaluation}
\label{sec:eval}

\subsection{Experimental Setup}
\label{sec:setup}

\textbf{Datasets.} The attack corpus comes from real carpet-bombing attacks captured by a security cloud vendor in 2023--2026: 71 files, 10 types, three forms (Table~\ref{tab:datasets}). The benign background includes CIC-IDS-2017 and four IoT benign files (CIC-IoT-2023-Benign, -Benign1/2/3). One backbone trace (2026-07-05, 197.5M records / 175.5M IPv4 packets, 15 min) serves the scale-limit analysis (Section~\ref{sec:scale}). The 30 holdout files (Section~\ref{sec:real}) stand outside the parameter-design process. Prior evaluations use different data mixtures: NetRadar uses CIC-IDS-2017/2018 benign traffic with real-world and simulated attack traces; DoLLM trains on CIC-DDoS2019 and a real ISP NetFlow trace; GMCB/CBDDoSCLIP use modified public datasets. Our 71 files are full-packet captures from 2023--2026, the largest such corpus we are aware of in this line.

\begin{table*}[t]
\centering
\caption{Attack corpus statistics.}
\label{tab:datasets}
\begin{tabular}{@{}P{4.2cm} r r r r r r@{}}
\toprule
Category & Files & Packets & \makecell[c]{Duration / span\\(s)} & \makecell[c]{Distinct\\destinations} & \makecell[c]{Median packets\\per destination} & \makecell[c]{Median packets\\per flow} \\
\midrule
ACK sweep & 6 & 42,069 & 156.0 & 770 & 23 & 1 \\
DNS reflection sweep & 1 & 5,000 & 0.0\textsuperscript{\textdagger} & 256 & 19 & 1 \\
HTTP sweep & 5 & 214,713 & 586.3 & 388 & 315.5 & 6 \\
HTTP reflection sweep & 2 & 188,063 & 427.4 & 106 & 1774.5 & 1 \\
HTTPS reflection sweep & 1 & 12,832 & 359.2 & 46 & 275.5 & 1 \\
ICMP sweep & 11 & 1,281,821 & 465.3 & 818 & 1077.5 & 367 \\
SYN sweep & 21 & 3,237,350 & 4037.7 & 1479 & 2112 & 1 \\
TCP reflection sweep & 1 & 99,947 & 283.4 & 73 & 1399 & 1 \\
UDP sweep & 20 & 1,419,309 & 178.5 & 1471 & 324 & 5.5 \\
WEB sweep & 3 & 117,639 & 694.2 & 285 & 295 & 7 \\
Real multi-prefix (2 natural files) & 2 & 962,959 & 600.0 & 502 & --- & --- \\
Synthetic multi-prefix, by timestamp (13 files) & 13 & 8,658,144 & 91,283.3 (span) & 175 & --- & --- \\
Synthetic multi-prefix, shuffled (15 files) & 15 & 9,621,103 & 694.2 & 196 & --- & --- \\
\bottomrule
\end{tabular}

\vspace{3pt}
{\footnotesize
\textsuperscript{\textdagger}This file lasts less than 0.05 s, so its duration is recorded as 0.0. The 91,283.3 s entry is the span of the concatenated timeline, not an attack duration. \\
The three real/synthetic multi-prefix categories total 30 files, 140 sample--victim segments, and 50 victim /24s; summary rows take the upper median of per-file values (the upper-median convention of Table~\ref{tab:params}; for $n=2$ this is the larger value).}
\end{table*}

\textbf{Evaluation methodology.} We report true-positive (tp), false-positive (fp), and false-negative (fn) counts, with $F_1 = 2\mathrm{tp}/(2\mathrm{tp}+\mathrm{fn}+\mathrm{fp})$. Attack and benign traffic never co-occur in any capture, so timelines are composed by the injection methodology of Appendix~\ref{app:data}; the holdout experiments rewrite no destination addresses; only the synthetic tiers rearrange timestamps. All results in this section come from a software implementation of the detector: the bucket state machine plus the control-plane confirmation, cooling, and cross-prefix logic of Section~\ref{sec:lifecycle}. Section~\ref{sec:plan} and Appendix~\ref{app:deploy-meth} describe the hardware forms and their relation to this implementation.

\textbf{Baselines and comparison conventions.} Table~\ref{tab:coverage} lists the ten baselines and where each is compared. The mixed-traffic experiment uses the destination-side spread estimators, and the low-rate and strategy experiments use the subset that produces a per-prefix decision online. NetRadar is the supervised external comparison.

\begin{table*}[t]
\centering
\caption{Baselines per experiment, with exclusion reasons.}
\label{tab:coverage}
\footnotesize
\setlength{\tabcolsep}{3pt}
\begin{tabular}{@{}P{4.7cm} c c c P{6.8cm}@{}}
\toprule
Baselines & Static & Mixed & Deadline & Why not in the deadline comparison \\
\midrule
Entropy, MRB, Lakhina & \checkmark & \checkmark/\checkmark/-- & Entropy & MRB/Lakhina: represented by the entropy baseline (per-epoch; Lakhina also unbounded; Appendix~\ref{app:entropy}) \\
RHHH, ElasticSketch, HeavyKeeper & \checkmark & -- & -- & volume/heavy-hitter signal, not uniform sweep \\
TRW, Couper, SpreadSketch, SegSketch & \checkmark & -- & -- & source-side instantiations; source state unavailable under spoofing \\
NetRadar & -- & -- & \checkmark & supervised 1 s detector; external comparison in Sections~\ref{sec:netradar} and~\ref{sec:lowrate} \\
\bottomrule
\end{tabular}
\end{table*}

\subsection{Static Detection}
\label{sec:static}

\begin{table*}[t]
\centering
\caption{Static detection at prefix granularity.}
\label{tab:static}
\begin{tabular}{@{}l c r r r r r r r@{}}
\toprule
Method & Anchor\textsuperscript{\textdagger} & Memory (KB) & tp & fp & fn & Precision & Recall & F1 \\
\midrule
\textbf{SweepSketch (CIC)} & /24 & 512 & 55 & 1\textsuperscript{\textdaggerdbl} & 0 & \textbf{0.9821} & \textbf{1.0} & \textbf{0.991} \\
\textbf{SweepSketch (4 IoT datasets)} & /24 & 512 & 55 & 1\textsuperscript{\textdaggerdbl} & 0 & \textbf{0.9821} & \textbf{1.0} & \textbf{0.991} \\
Entropy (Lall'06) & /24 & 495.6 & 54 & 4 & 1 & 0.931 & 0.982 & 0.956 \\
Lakhina'05\textsuperscript{\S} & /24 & 136.9 & 52 & 3 & 3 & 0.945 & 0.945 & 0.945 \\
MRB & /24 & 509.7 & 54 & 9 & 1 & 0.857 & 0.982 & 0.915 \\
RHHH & /24 & 512.0 & 35 & 5 & 20 & 0.875 & 0.636 & 0.737 \\
ElasticSketch & /24 & 512.0 & 34 & 5 & 21 & 0.872 & 0.618 & 0.723 \\
HeavyKeeper & /24 & 512.0 & 34 & 5 & 21 & 0.872 & 0.618 & 0.723 \\
TRW & Source IP & 512.0 & 62,896 & 2,640 & 2,464,051 & 0.96 & 0.025 & 0.049 \\
Couper\textsuperscript{\P} & Source IP & 648.5 & 5,825 & 0 & 2,521,122 & 1.0 & 0.002 & 0.005 \\
SpreadSketch & Source IP & 511.9 & 3,773 & 37 & 2,523,174 & 0.99 & 0.001 & 0.003 \\
SegSketch & Source IP & 510.5 & 943 & 11 & 2,526,004 & 0.988 & 0.0 & 0.001 \\
\bottomrule
\end{tabular}

\vspace{3pt}
{\footnotesize
\textsuperscript{\textdagger}GT is all source IPs in the file for the four source-side instantiations, and the /24 prefix for the destination-side methods. All four are instantiated at the source anchor, including Couper, a generic multi-flow cardinality estimator (Appendix~\ref{app:baselines}). \\
\textsuperscript{\textdaggerdbl}The single FP prefix is CIC's local subnet 192.168.10.0/24, not an injection artifact: its per-window distinct count (p50 8, p95 12) meets the D2 spread floor, and legitimate volume bursts trigger the volume channel. Section~\ref{sec:benign} quantifies the resulting episodes. \\
\textsuperscript{\S}Lakhina reproduction: memory grows unbounded with traffic (136.9 KB is the state size on this dataset), so it appears only in this table; see Appendix~\ref{app:baselines} for the full convention. \\
\textsuperscript{\P}Couper provides official software and hardware implementations; we re-implemented it at its 648.5 KB configuration, the account used in the static row (Appendix~\ref{app:baselines}).}
\end{table*}

F1 is 0.991 on all five datasets, with one false-positive /24 on every dataset; the same-tier entropy baseline \cite{b12} scores 0.956, Lakhina \cite{b25} 0.945, MRB \cite{b26} 0.915. That gap is not significant: a file-level bootstrap gives overlapping 95\% intervals (0.958--1.0 and 0.897--0.987) and McNemar on the 55 GT /24s gives $p=0.50$; RHHH and ElasticSketch are significantly worse ($p<10^{-6}$); HeavyKeeper is also lower, at 0.723. Prefix-level recall is 1.0 for each of the 10 attack types. \textbf{Source-spoofing immunity}: after rewriting every packet's attack source to an independent address, SweepSketch's tp/fp/fn and F1 are unchanged (55/1/0, 0.991), source-agnostic by construction; the source-side baselines' detections drop to near zero. \textbf{Memory grid}: at 32 KB SweepSketch scores 0.963, against entropy's 0.339 at its smallest usable tier (45 KB). At 128 KB it scores 0.9818, nearly tying entropy's 0.9820 at 1013.7 KB with one-eighth the memory. From 256 KB onward it scores 0.991, above every baseline at every tier; volume-based methods (ElasticSketch/HeavyKeeper 0.72, RHHH 0.74) and source-side methods ($\le 0.096$) remain lower throughout. A per-/24 exact counter with the same gates reaches F1 0.982 at 3.7 MB (11.8 GB as a hardware bitmap for 5,767 distinct /24 keys (5,712 CIC + 55 attack GT)); SweepSketch reaches 0.991 at 512 KB.

\subsection{Co-occurring Benign and Attack Traffic}
\label{sec:mixed}

The mixed trace removes two artifacts of the injection design: reused target prefixes and injected timelines. That design reported F1 0.9008 and a 522 ms median onset offset (186--627 ms across the four IoT backgrounds). We replay the full CIC benign timeline four times (4$\times$8.09 h) and place all 71 real attacks into it in their original duration and rate, without rewriting destinations or timestamps; each attack appears once, and each round hosts a disjoint subset. The merged stream is 53,118,227 packets per seed, and we use five seeds.

\emph{Detection holds, and the control plane removes the false-alarm flood.} On the mixed stream the detector reaches tp 55, fp 1, fn 0 (precision 0.9821, recall 1.0, F1 0.991), identical across all five seeds; 61--66 of the 71 files (file-level) alarm within their attack window; victim-/24 recall over the full trace is 1.0 (55/55 GT /24s). Alarm-level precision is 0.893 over L1 episodes (0.869 including DIST), up from 0.27 without control-plane confirmation: confirmation, cooling, and quarantine reduce 4,927 raw dispersion/volume alarms (plus 11,732 twelve-destination signals) to 109 true + 13 false L1 episodes, with 19--21 quarantines. Entropy reaches F1 0.9550 and MRB 0.9076 on the same stream, a 3.6-point lead. The onset offset is p50 0.033--0.124 s with a tail to 1,169 s, the cost of per-packet decisions on sparse attacks. For the single-window detector, F1 rises from 0.9008 (injection) to 0.9735 (mixed); confirmation then lifts it to the reported F1.

\subsection{Direct Comparison with NetRadar}
\label{sec:netradar}

We use NetRadar's official feature extractor and its PointNet training/inference semantics on the same 106 packet captures, extracting 249,887 one-second windows. The captures contain no server telemetry, so we reconstruct the two URL-derived server-side features from HTTP payload where available and zero them otherwise; this is the same variant trained with random erasing. Reconstructed features leave the prefix-level F1 unchanged in all 25 runs; the comparison is a same-data reimplementation of NetRadar.

A five-fold victim-/24 split (same 10 attack types, unseen victim prefixes) is the comparison protocol. Over five training seeds and five folds, NetRadar reaches mean F1 0.9341 (bootstrap 95\% CI 0.8854--0.9674; 0.9525 under its /24 rule of at least eight hosts per epoch) and SweepSketch 0.9520; the difference is $-0.0179$ with a 95\% CI of $-0.0666$ to $+0.0154$, which includes zero. Accuracy is at parity; the two differences are cost (a 141 MB trained model versus 512 KB and no training) and latency (a one-second feature window versus a per-packet decision).

A leave-one-type-out control (Appendix~\ref{app:loto}) holds out one attack type at a time: NetRadar's mean F1 drops to 0.636 (recall 0.900; 9/10 types reach recall 1.0, DNS 0/1), while SweepSketch, which does not train, recalls 1.0 on all 10 types. On NetRadar's real-world attack trace (32 servers in one /24, 8.9M SYN packets, 3 s), SweepSketch flags 192.168.0.0/24 with the first D1 alarm 1.0 s after the first packet; this is a latency check on an external capture (Appendix~\ref{app:loto}).

\subsection{Benign Traffic and False Alarms}
\label{sec:benign}

\begin{figure}[t]
\centering
\includegraphics[width=\columnwidth]{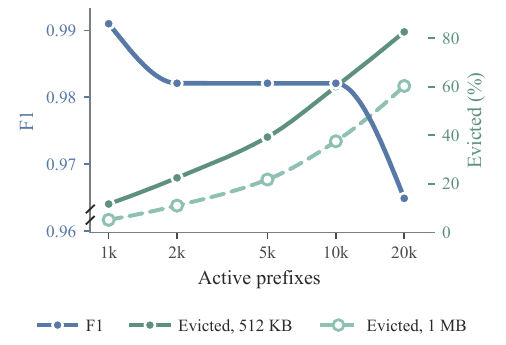}
\caption{F1 (left axis; axis starts at 0.96) and active-/24 eviction (right axis) as the number of active /24s grows; 18.24M packets per point.}
\label{fig:scale}
\end{figure}

We evaluated the detector on the full benign corpora. On CIC (11.6M packets, 8.09 h, 5,712 active /24s), the gates reduce the alarmed-/24 count from 7 (gates off) to 1 (gates on), and confirmation, cooling, and quarantine turn 1,164 raw alarms into 44 emitted L1 alarms, i.e., 10 episodes on 192.168.10.0/24 (1.24/h). All raw alarms fall on that prefix (685 D1, 377 D2, 102 twelve-destination); its per-window distinct count (p50 8, p95 12) meets the D2 floor, and legitimate bursts (up to 3,346 packets/window against a 102--510 EWMA baseline) trigger D2. The cross-prefix DIST channel adds zero false positives on CIC and IoT with gates on; the gates-off CIC run produces nine DIST alarms from shared /16 hits, suppressed by episode isolation. On the four IoT files (10.7M packets, 28 h) the gate-on configuration reports 38 episodes on one prefix (1.36/h); the gates-off run coalesces 24,412 raw D1 alarms into 12 episodes, while the gates-on run emits 392 raw alarms in 38 episodes; the prefix is the only residual, and aging re-arms the CUSUM. On a synthetic CDN-like workload (4 /24s, five seeds, three rates, 60 prefix-tests) all 60 prefixes alarm, but every alarm lies in a 60 s cold cut, one of the four cold episodes in each 1 h run; a newly seen CDN prefix needs a warm-up grace period. At 512 KB the detector still detects all 55 attack /24s at 20,000 active /24s (F1 0.9649; 82.6\% of active /24s are evicted over the run), while 1 MB leaves F1 unchanged and lowers eviction to 60.3\%; Figure~\ref{fig:scale} shows both trends.

\subsection{Low-Rate Short Attacks}
\label{sec:lowrate}

\begin{figure}[t]
\centering
\includegraphics[width=\columnwidth]{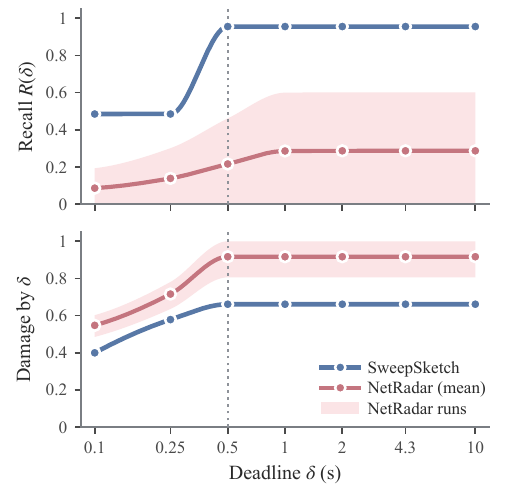}
\caption{Recall (top) and damage (bottom) against the alarm deadline $\delta$ (0.5 s attacks; 512 KB); the dashed line marks the attack end.}
\label{fig:deadline}
\end{figure}

We generate 0.5 s attacks with 32 destinations, rotate the target /24 every 0.5 s, run them on the busiest 600 s of CIC at 40 and 400 pps under three false-alarm budgets (0.1, 1, 5 episodes/prefix/day) and the same 512 KB. Across the grid (40/400 pps $\times$ $S=16/32$), SweepSketch alarms before the attack ends on 95.6\% of attacks at the 0.1 budget. At 40 pps/$S=16$ the cell rate is 93.0\% (p50 0.32 s; damage 80.1\%); at 400 pps/$S=32$, 98.7\% (p50 0.03 s; damage 39.8\%). At the 0.1 budget NetRadar's two-seed mean is 14.8\%; both seeds miss every 40 pps/$S=16$ attack; seed 1 reaches 87\% at 400 pps/$S=32$ under the 1/prefix/day budget. Figure~\ref{fig:deadline} plots the curve; Figure~\ref{fig:realonset} reports the real-file onset: 47.9\% alarm within 0.5 s and 49.3\% within 1 s; among the 40 files reaching 12 destinations within 0.5 s, 85\% alarm by then (median 0.0 s), whereas NetRadar reaches 22.5\% and 0\% with a one-second dwell. The window-batch estimator alarms on 0/213 file--deadline trials (Appendix~\ref{app:entropy}).

\textbf{Proposition T7 (latency--damage floor).} Consider a uniform sweep over $S$ destinations at per-prefix rate $\lambda$. Any detector whose first alarm requires $k$ new destinations has latency $\ge k/\lambda$ and damage-at-alarm $\ge k/S$; a window-batch detector evaluates only at boundaries, so its latency is at least the distance to the next boundary and can exceed the duration of a sub-window attack. SweepSketch's quiet-floor gate uses $k=12$; its observed 0.32 s at 40 pps matches $12/40=0.30$ s, and observed damage is 80.1\% at $S=16$ and 39.8\% at $S=32$, above the 75\% and 37.5\% floors. Measured damage is 39.8--80.1\%, against 82--100\% for window-batch alternatives.

\subsection{Deadline-Recall Curves}
\label{sec:deadline}

\begin{figure}[t]
\centering
\includegraphics[width=\columnwidth]{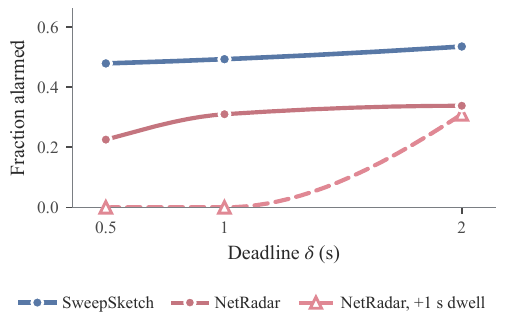}
\caption{Fraction of the 71 real captures alarmed by deadline $\delta$; also shown are NetRadar and its one-second-dwell variant.}
\label{fig:realonset}
\end{figure}

Recall is the fraction of attacks alarmed by $\delta$; damage is the fraction of victims covered by then. SweepSketch reaches its eventual recall by $\delta=0.5$ s; NetRadar reaches 21.6\% at that deadline (averaged over its three false-alarm budgets and two training seeds) and plateaus at 28.7\% by $\delta=2$ s. At the operational deadline (attack end, 0.5 s), damage is 66.1\% for SweepSketch and 91.7\% for NetRadar; the window-batch estimators are in Appendix~\ref{app:entropy}. The same pattern holds on the 71 real files (Figure~\ref{fig:realonset}). On the four unique sub-60 s real attacks, the median damage at first alarm is 0.0945, against 0.877 and 0.474 for entropy at its mixed-trace and operational thresholds; strict recall is 100\% against 50--75\%. On the remaining files, 94.4\% of real attacks last at least 60 s, where all detectors are fast enough.

\subsection{Validation on Real Attack Samples}
\label{sec:real}

On the 30 holdout files (2 real captures plus 28 deterministic re-orderings of 15 captures), SweepSketch detects all 50 victim /24s (140 sample--victim segments; fn=0, fp=0 within attack files), with first alarms at 0--5.4 s from cold start. The three tiers (2 natural files, 13 synthesized, 15 shuffled) are detected identically. Detection is insensitive to temporal structure; the 30 files are not 30 independent attacks, as they share 15 underlying captures. Volume-threshold baselines miss 27--30\% of these samples, entropy/MRB detect everything, and Lakhina detects 136/140; the four source-side methods recall 0.21--5.63\%.

\subsection{Ablation and Component Contributions}
\label{sec:robustness}

A 17-variant ablation leaves F1 within 0.01 of the base in 15 variants (14 bit-identical); the exceptions are $K_{\mathrm{FLOOD}}=4$ (0.9402) and the 16-register variant (0.973), so static detection is largely carried by the T-HLL core. The gates and the hierarchy act on temporal semantics instead: Table~\ref{tab:components} (Appendix~\ref{app:tables}) gives the component ablation and the 18-configuration one; the two channels are complementary.

\subsection{Adversarial Evaluation}
\label{sec:adversarial}

Under Kerckhoffs's assumption the attacker chooses the packet ordering to evade detection. A 192-cell strategy grid evaluates the spatial boundary ($S$ distinct destinations per window $\times$ $\lambda$ packets per window $\times$ ordering: scan, burst, and 70/30 split) under the injection mapping, so each cell isolates the rate, ordering, and density variables; collision loss under the production hash is handled separately by Lemma T5's $\delta(S)$.

The detected region is the union of the three spatial floors: scan cells with $S\ge12$ and $\lambda\ge16$; burst cells whose event density exceeds the 41\% drift threshold; and $\lambda\ge256$, $S\ge6$ cells detected by D2 in one window. All $S=4$ cells fall below $K_{\mathrm{FLOOD}}=6$; burst cells with $S\le16$ and $\lambda\in\{64,128\}$ stay below both floors; the 70/30-split $S=16$ cells need four windows because $n_{\mathrm{new}}=11$ fails the quiet floor. The fourth floor, dense-window absorption, is temporal and is measured in the injection experiment. Within the grid the boundary coincides with the floors; an adaptive search against the published floors is reported in Section~\ref{sec:attacker}.

\subsection{Adaptive Attacks and Cross-Prefix Correlation}
\label{sec:attacker}

This grid fixes a small set of orderings. We search 9,415 strategies (47,685 strategy--checkpoint rows) over $(S,\lambda,\text{ordering},\text{target rotation},\text{rate})$ under Kerckhoffs's assumption, with five seeds; the fast simulator is validated against the reference state machine on 3,124 windows with zero mismatches. A per-packet-rotation sweep is detected in its first window on 5/5 seeds.

\emph{Aging, confirmation, and cross-prefix correlation close the evasion surface.} A worst-case strategy uses three-packet bursts per destination and switches to a different /16 every window. It touches 25\%, 50\%, 75\%, and 90\% of a /24's hosts within 2, 4, 6, and 8 windows (8.6--34.4 s), using 2,048--8,192 packets, and the baseline produces 0/5 alarms. We therefore add bucket aging, confirmation with cooling, and cross-prefix correlation (Section~\ref{sec:lifecycle}).

\emph{With these components the attacker is detected inside the envelope.} For attacks above the published floors ($S\ge16$ and $\ge100$ pps), the full detector detects 100\% of strategies at the 50\%, 75\%, and 90\% damage checkpoints (330 union-damage strategies; 113 single-/24 strategies at 50\% damage and 84 at each of 75\% and 90\%, five seeds each) and 95.1--96.2\% at 25\% damage, with the /16-switching worst case detected in window 2 (4.3--8.6 s after onset) on 5/5 seeds. Confirmation removes 236 one-off detections entirely (176 at the /24 level and 60 at the /16 level). None of these 236 removals is in-envelope at 50--90\% damage (all have $S\le11$); at 25\% damage, confirmation costs 4.8\% of in-envelope strategies, because a 400 pps burst reaches the 25\% checkpoint in window 2 while confirmation needs window 3. The residual evasions are exactly the published floors: $S<16$ at low rate, and a single /24 below 47 pps, where both configurations stay silent by design.

\emph{Benign cost.} On CIC the benign cost is unchanged (Section~\ref{sec:benign}): one benign prefix is alarmed and three are quarantined. A separate target-hopping experiment switches the target /24 every one or two windows, with per-packet rotation inside each window. The full detector reaches 88.7--89.8\% victim-weighted detection at dwell 1 and 94.4--95.3\% event-weighted detection at dwell 2 (84.1--87.8\% and 90.9--92.8\% without confirmation), with first alarms at p50 0.03--0.35 s. Detection is zero for $S<12$ and 84--87\% for $S\ge16$; narrow sweeps, not the switching rate, define the residual blind spot. Confirmation also reduces false events in that experiment from 28 to 3.
\section{Deployment on Programmable Hardware}
\label{sec:deploy}

\subsection{Deployment Forms and Control-Plane Split}
\label{sec:plan}

We instantiate one decision function in three forms: a server reference, a Tofino2 data-plane program, and a commercial DPU micro-engine. Table~\ref{tab:forms} summarizes where each form keeps the per-packet state and where it evaluates the final comparison. 

In the switch form, we compile, place, and deploy the complete per-packet state chain on a Tofino2 switch (Tofino2 Native Architecture, T2NA, bf-p4c 9.7.1): 90 physical tables (89 algorithmic plus the shared IPv4 longest-prefix-match (LPM) table), 12 stages, and 13 stateful registers each confined to a single match-action-unit (MAU) stage, with zero placement errors and 75\% peak packet-header-vector (PHV) occupancy. The register state is 500 KB, inside the 512 KB ceiling. Only the shared IPv4-LPM table uses TCAM; the algorithmic structure uses none (Table~\ref{tab:p4}). The program reports a 28 B state snapshot per packet (a \emph{digest} is the alarm summary); the algorithmic bucket, from which that state is packed, is 44 B.

The program is a two-pass resubmit pipeline: pass 1 reads the bucket state and computes the per-packet event; pass 2 writes the new state. The switch keeps the per-packet state chain (bucket key, epoch, T-HLL array, CUSUM accumulator, and the counters) and reports a state snapshot on every packet. It also evaluates the per-packet terms of the gate predicates ($pkt \ge P_{\mathrm{MIN}}$, $n_{\mathrm{new}} \ge 2\,n_{\mathrm{ewma}}$) and carries the held candidate in the snapshot. In this form the control plane maintains the per-window baselines, completes the two gate predicates on each snapshot, and runs confirmation and cross-prefix correlation. The split changes placement only: both predicates and the per-packet decision are unchanged, and because every packet produces a snapshot it adds no window of latency. Section~\ref{sec:throughput} measures the control-plane CPU it adds; Appendix~\ref{app:deploy-meth} gives the register-placement reason and the field-by-field division.

\begin{table}[b]
\centering
\caption{Deployment forms and where each keeps the per-packet state and the final comparison.}
\label{tab:forms}
\footnotesize
\setlength{\tabcolsep}{3pt}
\begin{tabular}{@{}P{1.9cm}P{2.1cm}P{1.6cm}P{2.0cm}@{}}
\toprule
Form & Per-packet state (budget) & Final comparison & Evidence \\
\midrule
Server reference & bucket, 512 KB & bucket & C++, 13.2 Mpps/core \\
Tofino2 & switch registers, 500 KB & control plane & 90 tables, 12 stages \\
DPU micro-engine & on-chip SRAM, 240 KB & microcode & 5,120 slots, 1,135 instructions \\
\bottomrule
\end{tabular}
\end{table}

In the micro-engine form, a match-action stage classifies protected traffic and forwards it to a pool of general-purpose RISC-V micro-engines. The per-packet program reads the destination address, fetches the global epoch, updates the /24 and /16 buckets in on-chip SRAM, evaluates both gates, maintains the held bit, and appends a 32-byte alarm record to a lock-free ring. The control plane publishes the epoch, drains the ring, and runs confirmation and cross-prefix correlation. The state is 240 KB, 5,120 slots of 48 B (4,096 /24 and 1,024 /16); the microcode is 1,135 instructions, 4.6 KB of a 32 KB instruction memory, and a packet costs 13 on-chip memory round trips. The instruction set has no multiplier, so the hash is the multiply-free lookup3 that the reference state machine also uses. The micro-engine is a general-purpose integer core, so the complete decision chain runs on the fast path. We compile the microcode with the vendor RISC-V toolchain and build the device image, which we deploy on the DPU. We verify the decision logic bit-identically on 10.47 M packets against the reference (Appendix~\ref{app:deploy-meth}).

A TCAM-lookup variant of the same program moves 19 lookup tables to ternary memory: TCAM rises from 8 to 29 cells and match-table SRAM falls from 11,264 to 2,048 words, while the two-pass structure, the 12 stages, and the single-stage register attachment are unchanged (Table~\ref{tab:p4}). TCAM is read-only and cannot replace a register read-modify-write.

\subsection{Resource Accounting}
\label{sec:resources}

Table~\ref{tab:p4} gives the compiled resources of the two lookup forms; match-table SRAM and the 512 KB register account are separate accounts (Appendix~\ref{app:convb}).

\begin{table}[t]
\centering
\caption{SweepSketch deployment resources on Tofino2 (bf-p4c 9.7.1).}
\label{tab:p4}
\footnotesize
\setlength{\tabcolsep}{3pt}
\begin{tabular}{@{}l r r@{}}
\toprule
& \makecell[c]{SweepSketch\\(SRAM)} & \makecell[c]{SweepSketch\\(TCAM)} \\
\midrule
Physical tables & 90 & 90 \\
Stages & 12 & 12 \\
Stateful ALU (registers) & 13 & 13 \\
TCAM (cells; share) & 8 (1.67\%) & 29 (6.04\%) \\
MAU SRAM (words) & 11,264 & 2,048 \\
SRAM (units) & 73 & 76 \\
Register state & 500 KB & 500 KB \\
Snapshot & 28 B & 28 B \\
\bottomrule
\end{tabular}

\vspace{2pt}
{\footnotesize
\textit{Note:} MAU SRAM words count match-table entries; SRAM units are compiler resource units, dominated by the 13 state registers. \\}
\end{table}

The micro-engine form trades the match-table account for a memory account: 240 KB of SRAM for 5,120 slots of 48 B, and 10 KB of slot locks. On the 71-capture corpus it reproduces the 512 KB account's raw alarm count exactly (594 dispersion and 240 volume alarms).

\subsection{Control-Plane Cost}
\label{sec:throughput}

The reference state machine runs at 141.7 kpps on one core (134.7--148.2 over three repeats, 9.5\% spread), and 108.4 kpps end-to-end on the full CIC trace. The optimized C++ implementation is functionally identical (zero differing alarms in five checks, including the full 11.6M-packet CIC trace at 44/44 emitted alarms and a synthetic path-coverage trace) and reaches 13.21 Mpps detect-only (13.14--13.56 over three repeats, p50 71.5 ns/packet) and 1.482 Mpps end-to-end with packet-capture parsing (1.415--1.498, p99 1.10 $\mu$s). In the switch form the control plane maintains the per-window baselines, evaluates the final gates, and runs confirmation and alarm issuance. At 5k--20k active /24s and 47--100 pps each, it receives 0.2--2.0 Mpps of 28-B snapshots (5.6--56 MB/s); the C++ implementation, which also runs the full state machine, needs one core up to 1.4 Mpps and two at 2.0 Mpps in its end-to-end configuration, so this is a conservative bound on the control-plane work. In the micro-engine form the control plane receives one 32-byte record per alarm and a 4.29 s epoch tick, not a snapshot per packet, so its load is bounded by the alarm rate.

\section{Discussion}
\label{sec:discussion}

\subsection{Deployment Scale Limits}
\label{sec:scale}

For the deployment setting of Section~\ref{sec:threat}, the envelope is 5k--20k active /24s at 47--100 pps each (Section~\ref{sec:throughput}; not a bound on how many /24s can be swept at once). On the backbone trace (Section~\ref{sec:setup}), 4.71M active /24s share 7,943 bucket slots (593$\times$ over the 15-min trace, i.e., time reuse rather than simultaneous occupancy); 2,470 benign /24s cross $n_{\mathrm{new}}\ge12$ in at least one window, and 2,481 /24s raise at least one L1 alarm (the volume channel admits $n_{\mathrm{new}}\ge6$, which accounts for the difference). The benign-only run yields 9,323 episodes and 716 DIST alarms. At 512 KB, 63/71 attacks are detected (F1 0.0482); at 8 MB, 62/71 (F1 0.0454). Eight of the 71 attacks are missed at 512 KB through same-key reuse; at 8 MB a ninth miss is the four-window aging grace, which also explains the one-file difference between the budgets. The backbone trace is an out-of-envelope stress test.

\begin{figure}[t]
\centering
\includegraphics[width=\columnwidth]{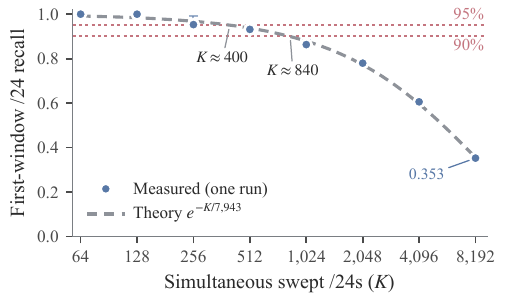}
\caption{First-window /24 recall against simultaneously swept /24s $K$ (200 pps each, one per /16; reference state machine). Points: one run per $K$; bars: five-seed min--max at four $K$; dashed: $e^{-K/7,943}$.}
\label{fig:capacity}
\end{figure}

A separate simulation of the reference state machine sweeps $K$ /24s at an equal rate of 200 pps over eight windows across 7,943 /24 slots. First-window recall is close to $e^{-K/7,943}$ (dashed in Figure~\ref{fig:capacity}), which reaches 95\% at $K\approx400$ and 90\% at $K\approx840$; misses are exactly the keys that collide in a bucket, and a detected key alarms in the first window (p50 60 ms, 4.69\% damage). The /16 level has half the slots (3,972), and when each swept /24 lies in a different /16 it sees the same $K$ keys and saturates earlier (95\% at $K\approx205$; 0.121 at $K=8,192$). Its role is to aggregate staggered attacks, where several /24s share a /16; the two levels fail independently. The micro-engine form trades slot count for fast-path completeness: its 5,120 slots are half the /24 count of the 512 KB account, so the concurrent-prefix capacity curve shifts left by a factor of 1.9; on the 71-capture corpus the two accounts produce the same alarms.

\subsection{Detection Floors and Their Empirical Counterparts}
\label{sec:lowerbounds}

Lemma T6's four floors each have an empirical counterpart in the strategy grid (Section~\ref{sec:adversarial}). Dense-window absorption drives all 12 injection misses: earlier same-target windows raise the EWMA baseline to $n_{\mathrm{ewma}}=11$--$38$, so the change gate stays above $n_{\mathrm{new}}$ (per-window states in Appendix~\ref{app:tables}). The tag-wraparound case was measured with hosts returning every eight windows.

\subsection{Where the Decision Chain Meets the ISA}
\label{sec:isa}

The switch form's remaining decision elements (the packet-rate EWMA, the held bits, and the two alarm bits) do not fit the Tofino2 action model: an action admits at most two PHV sources (one when a constant or action data is present), and a 16-bit scratch value is split across containers, so a two-operand EWMA update can need up to five sources after allocation. A structural skeleton with 18 registers, a 17-byte resubmit header, and a 21-byte alarm record passes PHV allocation and placement at 79.4\% peak occupancy, but the gate logic falls outside the action-source model. The micro-engine form evaluates the same gates on a general-purpose integer core without PHV co-design; the Tofino2 constraint is therefore the action model.

\subsection{Limitations and Claim Boundaries}
\label{sec:limitations}

The 30 holdout files are not independent attacks (Section~\ref{sec:real}), and the victim-/24 split is the stricter protocol. Detection scope is carpet bombing with a uniform destination-IP distribution; single-target flooding belongs to host-level super-receiver detectors. Appendix~\ref{app:entropy} reports the window-batch estimators; the main external comparison is a reimplementation of NetRadar. D2 remains the residual channel (Section~\ref{sec:benign}). T-HLL's event count is also biased under the production hash and its 3-bit tag can miss periodic returns; Section~\ref{sec:thll} quantifies both, and T5's $\delta(S)$ accounts for the former. Both forms are deployed on hardware (a Tofino2 switch and a DPU); our testbed did not support live-traffic tests. We report compiler and placement results for the switch form, and compiler results plus a bit-identical host replay for the micro-engine form. The throughput figures are for the C++ implementation of the same state chain.

\section{Related Work}
\label{sec:related}

\textbf{Carpet-bombing detection.} Academic characterization begins with PAM'21 \cite{b9}. NetRadar \cite{b6} analyzes unsampled raw traffic per prefix and reports roughly 1 s detection latency. GMCB \cite{b7} uses a graph model, motivated by slow responses to minute-scale attacks. DoLLM \cite{b8} processes flow data with a large language model and reports that inference latency limits real-time effectiveness (survey: \cite{b30}); CBDDoSCLIP \cite{b24} is a lightweight multimodal framework. SweepSketch differs in mechanism: per-packet decisions in fixed-width memory.

\textbf{Data-plane measurement and sketches.} SegSketch \cite{b3} detects super hosts via segmented cardinality estimation; we use its exported counts under the same 512 KB accounting. SpreadSketch \cite{b4}, Couper \cite{b5}, RHHH \cite{b27}, ElasticSketch \cite{b28}, and HeavyKeeper \cite{b29} are likewise measurement sketches: they provide only counts and estimates, with no alarm semantics. TRW is the exception among the source-side baselines: it raises a SPRT alarm rather than reporting counts, and the static row uses the reference implementation's detections. Jaqen \cite{b31} is a system template for data-plane detect+respond (double buffering, on-chip signals, on-chip records), but its signal is per-victim traffic change, which a uniform sweep does not produce. Our D2 channel borrows its change-detection idea and replaces the per-victim signal with a /24-aggregated one.

\textbf{Change detection and sequential testing.} TRW \cite{b2} applied SPRT to port-scan detection (per-source connection success rate), the engineering precedent for sequential testing, but its source anchoring and connection-visibility assumptions fail under carpet bombing (Table~\ref{tab:static}/Section~\ref{sec:real}). Our CUSUM decision inherits the optimality of Wald \cite{b19} and Wald--Wolfowitz \cite{b13}. The quantization and truncation bounds come from \cite{b20}, \cite{b21}, and \cite{b22}.

\textbf{Theoretical lower bounds.} Sample and memory lower bounds for distribution testing (\cite{b15}, \cite{b16}) and streaming lower bounds for entropy estimation (\cite{b11}'s $\Omega(\varepsilon^{-2}/\log(\varepsilon^{-1}))$; \cite{b23}'s $k \cdot b = \Omega(n/\varepsilon^2)$) underpin the comparison in Section~\ref{sec:theory}. We contrast each one with the task difference (one-sided test with a fixed alternative versus two-sided tolerant testing) in Section~\ref{sec:lemmas} and Appendix~\ref{app:proofs}.

\section{Conclusion}
\label{sec:conclusion}

SweepSketch turns sketch state into damage-bounded alarms: it flags carpet bombing while the attack is still in progress, covers all holdout victim prefixes, and matches the supervised model on same-type splits without training. The state chain is placed in a Tofino2 switch, and the full decision chain fits a DPU micro-engine.

\appendices

\section{Notation and Derivations}
\label{app:notation}

\textbf{The record field, $\rho$, and the 44 B budget.} The register index $r$ takes the low five
bits of $H(\mathrm{dst})$; the 27 bits above it form the record field, whose leading-zero count is
$\rho$. With $H$ uniform, $\Pr(\rho \ge k) = 2^{-k}$: $\rho$ is a truncated geometric, so five bits
suffice. Two destinations tie with probability $\sum_{k \ge 0} 4^{-(k+1)} = 1/3$,
which is the $R_2$ of Lemma L1 and, by symmetry, $\Pr(\rho_2 > \rho_1) = 1/3$; this fixes L1's $2/3$
loss coefficient and the $1 - 7/96 \approx 0.927$ margin at $S = 8$ behind $\theta_1$. This split gives the layout: 44 B $=$ 32 B of registers $+$ three 32-bit words. Each 8-bit register holds a 5-bit $\rho$ and a 3-bit window tag; the 5-bit $r$ selects the register. The tag wraps after eight active windows (Section~\ref{sec:thll}). $d2_{\mathrm{held}}$ is a separate one-bit latch in the bucket metadata, not a bit of $C$.

\textbf{The 4.29 s window.} An epoch is the $2^{32}$ ns wrap of the ingress nanosecond timestamp;
bits $[36{:}32]$ give the five-bit epoch field, so one window is $2^{32}\,\mathrm{ns} = 4.294967296$ s; it sets only the counting and baseline-reset granularity; decisions stay per packet, hence the millisecond latencies of Section~\ref{sec:lowrate}.

\textbf{The threshold $h$.} $h$ is chosen so that a repetition-free sweep crosses it on its second
packet, i.e.\ $h \le 2z_+ = 76$; the frozen value is $h = \mathrm{round}(9.2 \times 8) = 74$ (9.2 is the continuous threshold and 8 the fixed-point scale), and T2
\emph{evaluates} it at $e^{-\gamma^* h} \le 1.8 \times 10^{-3}$ per window per bucket rather than
deriving it.

\section{Proof Details}
\label{app:proofs}

\textbf{Proof of Lemma L1.} (1) Contribution structure: the first touch of register $r$ always produces an event (tag mismatch); the remaining events are the strict-$\rho$ records among the $k_r - 1$ subsequent hits. The record process depends only on the order of $\rho$ values (exchangeable), so $\mathbb{E}[\#events \mid \{k_r\}] = \sum_r (1 + R_{k_r})$, where $R_k$ is the expectation of the geometric tie-breaking record process ($R_2 = \Pr(\rho_2 > \rho_1) = 1/3$; in the ideal continuous case $R_k \le H_k - 1$, and ties only reduce records). (2) Loss-free lower bound: for $k \ge 2$, $\mathbb{E}[\mathrm{loss}] = (k-1) - R_k \le (k-1) - 1/3 \le (2/3)\cdot\binom{k}{2}$ (equality at $k=2$; for $k\ge 3$, $k - 4/3 \le k(k-1)/3 \iff (k-2)^2 \ge 0$). Hence $\mathbb{E}[n_{\mathrm{new}} \mid S] \ge S - S(S-1)/96$. (3) Injection mapping: when $S \le 32$ and the mapping is injective, each register carries at most one distinct destination, and $n_{\mathrm{new}} = S$ pointwise.

\textbf{Proof of Lemma L2.} Under the injection mapping every packet fires, so the event fraction is 1. Without injection, when $S$ packets match $S$ distinct destinations, L1 gives $\mathbb{E}[n_{\mathrm{new}}\mid S]\ge S-S(S-1)/96$; dividing by $S$ gives $1-(S-1)/96$, which at $S=8$ is $1-7/96\approx0.927\ge\theta_1$.

\textbf{Proof of T1.} (a) Unbiasedness of the collision count: $\mathbb{E}[\sum_r \binom{k_r}{2}] = \binom{S}{2}/m$ (linearity of pairwise collisions), an unbiased estimator of $\|p\|_2^2$ (the fixed-width instance of \cite{b16} Thm 3.2.2). (b) The sample complexity $\Theta(\sqrt{m}/\varepsilon^2)$ for $m=32$ registers is cited directly from \cite{b15} (Thms 3/4). (c) Tagged self-cleaning as a coupon-process mapping: with fresh-$\rho$ and mismatch-as-event, each window's event process is isomorphic to a first-occupancy coupon-collector process; the per-window baseline reset (A1) gives independence across windows, and T-HLL is the fixed-width representable implementation of ``one coupon experiment per window''. (d) Implementation-parameter correspondence: $S_{\mathrm{min}} = 8$ is the minimum distinct count the decision needs.

\textbf{Proof of T2.} (a) Lundberg/Kingman bound (\cite{b17}): with $Z_t$ i.i.d., choose $\gamma^*$ such that $\mathbb{E}[e^{\gamma^* Z}] \le 1$ ($f(\gamma) = 0.0345e^{38\gamma} + 0.9655e^{-26\gamma}$, $f(0.0857) \approx 0.9997 \le 1$, converged by bisection in 60 rounds). $M_t = e^{\gamma^* C_t}$ satisfies the Kingman bound for reflected walks, $\mathbb{P}(\max_t C_t \ge h) \le e^{-\gamma^* h}$ (the reflection $\max(0,\cdot)$ does not increase the exceedance probability). (b) Wald's identity (\cite{b19}): the crossing stopping time $\tau$ satisfies $\mathbb{E}[C_\tau] = \mu_1 \mathbb{E}[\tau]$. Since $C_\tau \le h + z_+$, $\mathbb{E}[\tau] \le (h+z_+)/\mu_1$. (c) \cite{b18} (Azuma form): $|Z_t - \mu_1| \le 57.6 \le \Delta = 74$; here $57.6 = |z_-| + \mu_1 = 26 + 31.6$ is the one-step deviation bound, and $\Delta = 74$ is a conservative range parameter (the exact increment range is $z_+ - z_- = 64$). Also $C_N \ge T_N$ (reflection never decreases), and $T_N - N\mu_1$ is a martingale, so $\mathbb{P}(C_N \le h) \le \mathbb{P}(T_N \le h) \le \exp(-(N\mu_1-h)^2/(2N\Delta^2))$. (d) Quantization and truncation: 8-bit fixed-point quantization does not reduce the KL drift (\cite{b20}: $\mathbb{E}[\mathrm{LLR}]$ is non-decreasing). The average-run-length and expected-detection-delay (ARL/EDD) bounds for quantized CUSUM apply directly (Q-CUSUM, \cite{b21}). $C_{\mathrm{MAX}}$ saturation is a truncated SPRT (\cite{b22}: drift stays positive in saturation, and the OC/ASN (operating characteristic / average sample number) corrections are bounded). Numerical validation runs at a 1\% event rate, more skewed than $\theta_0$, where the negative drift is steeper and measured false positives are smaller. The direction of the bound relationships is unchanged.

\textbf{Proof of T3.} (a) ARL$_0$: the Kingman bound of T2-1 plus window independence (A1) give per-window alarm probability $\le \alpha_{\mathrm{walk}} \le e^{-\gamma^* h}$, so ARL$_0 = 1/\alpha_{\mathrm{walk}} \ge e^{\gamma^* h}$. (b) Latency: the Wald bound of T2-2 plus the $n_{\mathrm{new}}$ gate $G$ (both L1 branches, injection and collision). (c) Field-by-field bit-width accounting: T-HLL 256 bit (5-bit $\rho = \log(1/\varepsilon)$ scale, $\varepsilon = S_{\mathrm{min}}/256 \approx 1/32$), $C$ 16 bit ($h = 74$ plus saturation headroom, $\log(1/\alpha)/\gamma$ scale), $n_{\mathrm{new}}$/$n_{\mathrm{ewma}}$/$pkt$/$pkt_{\mathrm{ewma}}$ 8/8/16/16 bit (window state, counts bounded by saturation), and key+epoch+win 32 bit (bucket identity, invariant in $\varepsilon$/$\alpha$). (d) Lower-bound side: our task does not attain the Diakonikolas--Gouleakis--Kane--Rao (DGKR) bound \cite{b23} (the settings differ: DGKR is two-sided tolerant testing with error rate 1/3, $k$ samples $\times$ $b$-bit memory, $k \cdot b = \Omega(n/\varepsilon^2)$, strengthened to $\Omega(n\log n/\varepsilon^4)$ for $k < n^{9/10}$, $b \ge k^2/n^{0.9}$).

\textbf{Proof of T4.} (a) For a stable prefix, $n_{\mathrm{new}}$ has mean $\mathbb{E}[n_{\mathrm{new}}] \approx n_{\mathrm{ewma}}$ under $H_0$ (its Binomial mean with per-packet event probability $\theta_0$); the multiplicative Chernoff bound with factor 2 gives $\mathbb{P}(n_{\mathrm{new}} \ge 2n_{\mathrm{ewma}}) \le (e/4)^{n_{\mathrm{ewma}}} \le e^{-n_{\mathrm{ewma}}/3}$. (b) $d2_{\mathrm{held}}$: adjacent windows are independent (A1), so the joint false-positive probability is $3p^2-2p^3=O(p^2)$. (c) The real-valued recursion $x \leftarrow x + (n_{\mathrm{new}}-x)/8$ is a contraction mapping: $|x_t - n_{\mathrm{new}}| \le (7/8)^t |x_0 - n_{\mathrm{new}}|$ pointwise, and the absorption time $\min\{t : 2x_t > n_{\mathrm{new}}\} \le \log 2/\log(8/7) \approx 5.2$ windows (closed-form envelope: $2n_{\mathrm{new}}(1-(7/8)^t) > n_{\mathrm{new}} \iff t > 5.19$). The integer right-shift implementation satisfies $|x_t - n_{\mathrm{new}}| \le (7/8)^t |x_0 - n_{\mathrm{new}}| + 7$ ($d_{t+1} = d_t - \lfloor d_t/8 \rfloor \le 7d_t/8 + 7/8$ accumulates step by step); measured absorption is 7 windows (due to discretization and gate rounding).

\textbf{Proof of T5.} (a) The first touch is necessarily an event, so in the injection branch $n_{\mathrm{new}} = S$ pointwise (L1.3). (b) When $S \ge G$, $C$ crosses after the $\lceil h/z_+\rceil = 2$nd event of a repetition-free ordering (each event adds +38, a repeat subtracts 26), and the alarm latency is the $G$-th new-destination packet. (c) Hierarchical aggregation: the /16 level uses the same T-HLL structure, the same test, at a different resolution; the two levels' statistics are isomorphic, and their coverage union is every sweep with total distinct $\ge G$. (d) Reporting granularity is off by at most one level (on a /16 alarm, the active /24 buckets under that /16 are read on demand to localize the sweep). (e) Collision loss enters the detection-rate lower bound through the second-order term of L1.2: the single-window detection rate is $\ge 1 - \delta(S)$, where $\delta(S) = \Pr(n_{\mathrm{new}} < G \mid S)$ is the miss probability, computed exactly by combining the multinomial occupancy distribution (each register's occupancy has marginal $\mathrm{Bin}(S, 1/32)$) with the geometric tie-breaking record process ($\delta(12) \approx 0.77$, 10 windows 0.93; $\delta(14) \approx 0.23$, 3 windows 0.99; $\delta(8) \approx 0.45$); multi-window: $1 - \delta(S)^W$.

\textbf{Proof of T6.} Each floor follows directly from the inequality of its gate: $P_{\mathrm{MIN}}$/$S_{\mathrm{min}}$/$S_{\mathrm{quiet}}$/$K_{\mathrm{FLOOD}}$ are pointwise gates; the density threshold is the drift-sign condition $z_+p + z_-(1-p) > 0 \iff p > 26/64$; absorption is the fixed point of T4(c) (once the EWMA converges, the gate stays closed); the adversarial constructions map one-to-one onto the non-detected cells of the EXP8 grid and the fn composition of EXP4.

\textbf{Theory--implementation correspondence.} Table~\ref{tab:theoryimpl} maps every implementation parameter to its theorem and its Monte Carlo check (main-text pointer in Section~\ref{sec:params}).

\begin{table}[t]
\centering
\caption{Theory--implementation correspondence.}
\label{tab:theoryimpl}
\setlength{\tabcolsep}{4pt}
\begin{tabular}{@{}P{2.4cm}P{3.8cm}P{1.6cm}@{}}
\toprule
Implementation parameter & Theorem/Lemma & Monte Carlo \\
\midrule
$\theta_0$/$\theta_1 \to$ LLR $\{38,-26\}$ & Section~\ref{sec:cusum} derivation + T2 input & Pooled-sampling calibration + deterministic self-check \\
$h=74$ & T2 ARL bound ($e^{-\gamma^* h} \le 1.8\times10^{-3}$) & Monte Carlo suppression bound \\
$S_{\mathrm{min}}=8$, $S_{\mathrm{quiet}}=12$ & T5 (injection/collision branches) & Monte Carlo T5 + strategy grid \\
$\mathrm{EWMA\_SHIFT}=3$, $\mathrm{D2\_MULT}=3$, $d2_{\mathrm{held}}$ & T4 absorption / squared confirmation & Monte Carlo absorption curves \\
$P_{\mathrm{MIN}}=200$, $K_{\mathrm{FLOOD}}=6$ & T6 rate/over-concentration floors & strategy grid + injection misses \\
$C_{\mathrm{MAX}}=65535$ & T2 truncated SPRT & Drift stays positive in saturation \\
44 B/bucket, $N_{L1}/N_{L2}$ & T3 bit-width accounting & memory grid \\
\bottomrule
\end{tabular}
\end{table}

\section{Deployment Methodology}
\label{app:deploy-meth}

Table~\ref{tab:division} restates the division of labor of Section~\ref{sec:plan} for the compiled deployment.

\begin{table}[t]
\centering
\caption{Responsibilities of the data plane and the control plane in the switch form.}
\label{tab:division}
\setlength{\tabcolsep}{4pt}
\begin{tabular}{@{}P{2.1cm}P{1.7cm}P{3.9cm}@{}}
\toprule
Component & Deployment location & Responsibilities \\
\midrule
Per-packet state chain & Data plane & Bucket-index hashing, T-HLL event decision, lazy window rollover, CUSUM accumulator update, and window counters; a 28-B state snapshot per packet (key, $n_{\mathrm{new}}$, $C$, deltas, timestamp; pass 2 does not report) \\
Final decision & Control plane & Threshold gates (D1 dispersion, D2 volume), confirmation with cooling, and alarm issuance; the EWMA/held state is maintained per window \\
Alarm handling & Control plane & Cross-prefix correlation, digest aggregation and deduplication, operator alerts, and on-demand forensics \\
\bottomrule
\end{tabular}
\end{table}

In the micro-engine form the data plane additionally evaluates the final gates and maintains the held bit; the control plane retains confirmation, cooling, quarantine, and cross-prefix correlation.

\textbf{Why two passes.} Every access to a stateful register must map to one MAU stage, and the early read (old key, $C$, $pkt$) and the late write (the $C$ update after the HLL event) cannot share one; a 20-stage ingress budget does not remove the dependency. Pass 1 reads and decides, pass 2 writes; new values travel in an 88-bit resubmit header only on packets that update state. A single pass would move the cold/claim policy to the control plane or redesign the state machine.

\textbf{Micro-engine verification.} We compile the same C sources twice: once with the vendor RISC-V toolchain into the device image deployed on the DPU, and once for the host against a model of the on-chip memory, the per-slot locks, and the alarm ring. The host replay consumes the reference traces and dumps the raw alarm stream and the final bucket state; the reference state machine consumes the same traces. Over 73 traces (71 attack captures, 2.0 M benign packets, and a 1.85 M-packet mixed stream; 10,473,357 packets), all 13,659 raw alarms and all final bucket fields, including the 32-byte T-HLL array, are identical. The control plane, fed the same raw alarms, produces the same episode list as the reference on all 73 traces. On the 71-capture corpus the 240 KB form and the 512 KB account produce the same alarm counts (594 dispersion, 240 volume).

\textbf{Convention A collection.} Both forms receive the same IPv4-LPM forwarding table (dstAddr lpm, size 4096, covering port loopback); the SRAM form uses its 8 cells, and the TCAM form adds 21 cells from moving 19 exact-match lookup tables to ternary memory. TCAM shares use 24 cells per stage $\times$ 20 stages = 480; we count absolute cells rather than the compiler's average-stage column. The shared LPM is deployment scaffolding outside the 512 KB algorithmic budget.

\section{Datasets and Calibration}
\label{app:data}

\textbf{Composition of the 71 attacks.} See Table~\ref{tab:datasets}. \textbf{$K_{\mathrm{FLOOD}}$.} $K_{\mathrm{FLOOD}} = 6$ (D2's cold-start-window spread lower bound) is set at the boundary between flood-type attacks' per-window distinct counts (single digits to tens) and the benign median (about 2); the floors' theoretical structure is built by theorems T5/T6. \textbf{$\lambda_{\mathrm{min}}$.} The 40 pps design target is below the D2 volume floor ($\lambda_{\mathrm{min}} \approx 47$ pps; Table~\ref{tab:params}); the D1 dispersion channel is not subject to it. \textbf{Split discipline.} Calibration/test splits in Section~\ref{sec:params}; $\theta_0 = 0.0345$ (upper median of per-file medians over the 10-capture calibration corpus with 1-in-10 packet sampling, Table~\ref{tab:params}). \textbf{Holdout sample construction.} Natural multi-prefix samples: 2 files (8 adjacent /24s spanning a /21 boundary, with minimum-alignment covering prefix /20, and adjacent /23s; unprocessed). Timestamp-synthesized: 13 files (same-date files keep their real timestamps and are placed adjacently; across activities, concatenated in date order with 12 s non-overlapping gaps): real packets, real relative pacing, synthetic timelines concatenated in date order. Timestamp-shuffled: 15 files (t0-aligned; timestamps deterministically shuffled with a seed derived from the group name via md5; packet contents unchanged; timelines monotonically non-decreasing), as a synthetic-timeline control. Benign background: 2 files (CIC-IDS-2017 / CIC-IoT-2023). \textbf{Injection methodology.} Dynamic experiments compose timelines under strategy A (staggered injection onto distinct prefixes) and strategy B (remapping attack prefixes to idle prefixes). The holdout construction rewrites no destination addresses; only the synthetic tiers rearrange timestamps.

\section{Baseline Notes and Controls}
\label{app:baselines}

\textbf{Carryover of the static-table footnotes.} The Lakhina reproduction has no fixed memory budget, and we re-implemented Couper at 648.5 KB. The static table carries no AUC column: Couper's AUC of 1.0 next to its F1 of 0.005 is a source-anchor scoring distortion, not a strong method.

\subsection{Leave-One-Type-Out Control and NetRadar's Attack Trace}
\label{app:loto}

For the leave-one-type-out control, NetRadar reaches mean F1 0.636 (9/10 types recall 1.0; the per-type values are in Table~\ref{tab:loto}), while SweepSketch, which does not train, recalls 1.0 on all 10 types. NetRadar's prefix-split result therefore depends on same-type supervision.

NetRadar's real-world attack trace is a 3 s, $\approx$2.97 Mpps SYN flood against 32 servers in one /24 (8,896,996 packets; $\approx$92.7 kpps per server), captured at a cloud provider. It is high-rate and single-type, with no benign background and no server telemetry. Converting its packet records to pcap and running SweepSketch without retuning flags 192.168.0.0/24, with the first D1 alarm 1.0 s after the first packet and five L1 alarms; this is a detection/latency check, not an F1 measurement. It shows transfer to an external attack capture and a different measurement point, but does not replace the low-rate evaluation.

\begin{table}[h]
\centering
\caption{Leave-one-type-out results for NetRadar.}
\label{tab:loto}
\footnotesize
\setlength{\tabcolsep}{3pt}
\begin{tabular}{@{}l r r r r r r@{}}
\toprule
Held-out type & GT & TP & FN & FP & Recall & F1 \\
\midrule
ACK & 6 & 6 & 0 & 12 & 1.000 & 0.500 \\
DNS-Reflection & 1 & 0 & 1 & 1 & 0.000 & 0.000 \\
HTTP & 5 & 5 & 0 & 1 & 1.000 & 0.909 \\
HTTPS-Reflection & 1 & 1 & 0 & 9 & 1.000 & 0.182 \\
HTTP-Reflection & 2 & 2 & 0 & 1 & 1.000 & 0.800 \\
ICMP & 9 & 9 & 0 & 0 & 1.000 & 1.000 \\
SYN & 18 & 18 & 0 & 2 & 1.000 & 0.947 \\
TCP-Reflection & 1 & 1 & 0 & 3 & 1.000 & 0.400 \\
UDP & 20 & 20 & 0 & 2 & 1.000 & 0.952 \\
WEB & 3 & 3 & 0 & 3 & 1.000 & 0.667 \\
\bottomrule
\end{tabular}
\end{table}

\section{Resource-Accounting Conventions}
\label{app:convb}

\textbf{Compiled resource account.} The deployed program has 90 physical tables in the two-pass structure of Section~\ref{sec:plan}; the 13 state registers occupy 500 KB, and the program reports a 28-B snapshot per packet (Table~\ref{tab:p4}). Match-table memory is a separate account.

\textbf{Convention B (per-method byte accounts at each paper's minimal configuration).} SegSketch $\approx$103.5 KB; SpreadSketch 126 KB (paper's hardware tier) / 552 KB (1 MiB tier); Couper $\approx$0.5 MB (paper's minimal tier; its static row uses 648.5 KB); RHHH $\approx$32 KB; HeavyKeeper 10 KB ($k=100$); ElasticSketch 0.2 MB; MRB 96 KB; Entropy 32 KB; TRW has no tunable memory parameter. Table~\ref{tab:static} reports each method at its 512 KB configuration where one exists; for SegSketch these are 510.5 KB and 103.5 KB (the paper also reports a 32 KB tier), and for MRB 509.7 KB and 96 KB (31 KB tier). SpreadSketch uses the full 511.9 KB configuration ($w=1820$, folded 2,048-entry index). The entropy P4 row uses $z=64,g=3$; the memory-grid tiers below 64 KB use the reference schedule $g=\min(3,\max(1,z/4))$.

\textbf{Semantic mapping} (field-by-field correspondence between the reference implementation and the P4 registers). In the compiled form the data plane holds the bucket key, epoch, $n_{\mathrm{new}}$, $n_{\mathrm{ewma}}$ (read-only), $C$, $pkt$, and the $32\times8$-bit HLL array for both levels; the control plane receives the 28-B snapshot and completes the threshold gates, confirmation, and alarm issuance. Mapping: $epoch/win \leftrightarrow l1\_ep\{epoch[7{:}3],\ win[2{:}0]\}$; $n_{\mathrm{new}} \leftrightarrow l1\_nn$ (rollover reset, $+$event); $n_{\mathrm{ewma}} \leftrightarrow l1\_ne$ (read-only in the switch; the control plane maintains the baseline); $C \leftrightarrow l1\_C$ ($\pm llr$); $pkt \leftrightarrow l1\_pkt$; $hll[32] \leftrightarrow l1\_hll$ flattened $\{tag[7{:}5],\rho[4{:}0]\}$; bucket creation $\leftrightarrow$ claim (key written to $l1\_key$, win inherited and incremented, all other fields reset); $epoch\_of(ts) \leftrightarrow global\_epoch = tstamp[36{:}32]$; alarm deduplication $\leftrightarrow$ aggregation by (level, prefix, epoch). The second level uses the same mapping with $l2\_*$ registers; $l2$ has no $n_{\mathrm{ewma}}$ register, and its baseline is maintained on the control plane. The micro-engine form maps the same fields to a 48 B slot: a 16 B header holding the key, epoch, window, $n_{\mathrm{new}}$, $n_{\mathrm{ewma}}$, $C$, $pkt$, $pkt_{\mathrm{ewma}}$, and the held bit, followed by the 32 B T-HLL array; the gate predicates are evaluated in the microcode, and the control plane retains confirmation and correlation.

\section{Window-Batch Baselines}
\label{app:entropy}

We include the bounded-memory entropy estimator \cite{b12} as the representative window-batch baseline. At 512 KB, the entropy P4 convention gives 2,048 /24 slots of $z=64$ 32-bit counters. On the 0.5 s attack grid, a window-batch estimator can only evaluate at the 4.29 s boundary: at the 0.1 false-positive budget, entropy reaches 9.1\% strict recall (97.6\% within one epoch) with a 2.33 s median first alarm. On 71 real files it alarms within 0.5/1/2 s on 0/213 file--deadline trials.

\section{Reproduction Details}
\label{app:tables}

\textbf{Reproduction index.} EXP1--EXP9 correspond to static detection, source spoofing, the memory grid, dynamic detection, alarm flags, ablation, component toggles, the adversarial grid, and backbone scale; across all 9,415 strategies, including those below the floors, the mean detection rate is 0.54--0.57.

\textbf{Component contributions} (Table~\ref{tab:components}; main-text summary in Section~\ref{sec:robustness}):

\begin{table}[t]
\centering
\caption{Component ablation (synthetic timelines).}
\label{tab:components}
\setlength{\tabcolsep}{4pt}
\begin{tabular}{@{}P{1.6cm}P{3.4cm}P{2.6cm}@{}}
\toprule
Component & With component & Without component \\
\midrule
EWMA gates & Alarms converge in 7 windows (epochs 0--6, consistent with absorption theory) & 16 windows (epochs 0--15, sustained false positives) \\
Two-level hierarchy & The /16 level detects staggered attacks & Zero /16-level detections \\
Dual channels & Detects both scans and floods & D1 alone misses flood-type attacks; D2 alone misses scan-type attacks \\
\bottomrule
\end{tabular}
\end{table}

\textbf{Full ablation table} (18 configurations including base, F1; variant codes: m\# = register count; l1only/d1only/d2only = single level / D1 only / D2 only; nogate = EWMA gates removed; theta\_\# = $\theta_0$ variant; h\#/k\#/s\#/kf\#/pmin\# = CUSUM threshold / alarm gate / dispersion floor / cold-start spread bound / rate floor): base 0.991; m16 0.973 / m64 0.991 / l1only 0.991 / d1only 0.991 / d2only 0.991 / nogate 0.991 / theta\_005 0.991 / theta\_05 0.9818 / h6 0.991 / h12 0.991 / k12 0.991 / kf4 0.9402 / kf8 0.991 / m2 0.991 / m4 0.991 / pmin100 0.991 / pmin400 0.991.

\textbf{Full memory-grid table} (F1): SweepSketch 32\slash64\slash128\slash256\slash512\slash1024 KB $=$ 0.963\slash0.972\slash0.9818\slash0.991\slash0.991\slash0.991; Entropy 45.1\slash112.6\slash247.8\slash495.6\slash1013.7 KB $=$ 0.339\slash0.922\slash0.916\slash0.956\slash0.982; MRB 31--1022 KB $=$ 0.900--0.915; RHHH 0.737, ElasticSketch/HeavyKeeper 0.646--0.723, TRW 0.003--0.096, Couper 0.0--0.009, SpreadSketch 0.0--0.006, SegSketch 0.0--0.002 (all tiers).

\textbf{C++ insertion-path throughput.} Table~\ref{tab:throughput} compares the C++ insertion paths of the baselines and SweepSketch; the detector's own end-to-end throughput is in Section~\ref{sec:throughput}.

\begin{table}[t]
\centering
\caption{C++ insertion-path throughput (Mpps; 512 KB tier); full-detector throughput is in Section~\ref{sec:throughput}.}
\label{tab:throughput}
\setlength{\tabcolsep}{4pt}
\begin{tabular}{@{}l r r r@{}}
\toprule
Method & \makecell[c]{Backbone\\(164K /24)} & \makecell[c]{Lab\\CIC} & \makecell[c]{Attack\\(UDP sweep)} \\
\midrule
Lakhina & 104.8 & 503.5 & 530.0 \\
TRW & 56.1 & 500.9 & 52.1 \\
ElasticSketch & 109.1 & 304.7 & 340.7 \\
MRB & 183.0 & 220.4 & 221.5 \\
RHHH & 125.5 & 191.2 & 225.4 \\
Entropy & 299.0 & 164.9 & 167.1 \\
\textbf{SweepSketch} & \textbf{37.2} & \textbf{79.4} & \textbf{83.5} \\
Couper & 45.9 & 75.0 & 52.7 \\
HeavyKeeper & 43.5 & 63.4 & 73.9 \\
SpreadSketch & 43.6 & 54.0 & 37.8 \\
SegSketch & 21.5 & 31.3 & 19.3 \\
\bottomrule
\end{tabular}
\end{table}

\textbf{Throughput memory grid} (lab traffic, Mpps): SweepSketch 78.9--79.6 across all 32--1024 KB tiers; SegSketch 31.7--34.6; SpreadSketch 52.8--54.0; Couper 74.2--75.2; RHHH 172--205; ElasticSketch 174--311; HeavyKeeper 62.0--63.9; TRW 499--503; MRB 220--229; Entropy 370.7 at the 32 KB tier (whose $z = 6$, $g = 1$ configuration differs from the other tiers) and 164.6--167.2 from 64 KB up; Lakhina 515--525. (Independent timing rounds; same-tier values differ by $\le$4\%.)

\textbf{Per-window bucket states of the injection misses}: all 12 are dense-window absorption, with the change gate above $n_{\mathrm{new}}$ throughout (5) or no crossing co-occurring with $C \ge 74$ (7).

\textbf{Backbone scale table} (basis of Section~\ref{sec:scale}): on the backbone trace, the 8 MB pass raises 2,598 /24s, and 2,470 benign /24s reach per-window $n_{\mathrm{new}}\ge12$ (2,163 with $\ge16$, 1,533 with $\ge20$). The eight 512 KB misses are consecutive files that target the same /24 12 s apart, so the second file inherits the first file's EWMA state; at 8 MB the ninth miss is 43.248.130.0/24, the four-window aging grace. \textbf{Spoofing-variant details for source-side methods.} Per-file spoofed/unspoofed pairs; after spoofing, the source-side methods' detections drop to near zero (TRW: a single source; the released artifact uses only type-level aggregation ratios).\section*{Ethics Considerations}

The real attack traffic used in this paper was captured by a security cloud vendor within its authorized monitoring scope and is used solely for algorithm evaluation; the vendor-released form of each file is one row per packet with the fields \texttt{ts, src, dst, proto, sport, dport}. We do not interact with attack infrastructure in any way, and we provide no attack-enhancement tools; all experiments are offline traffic replay. The benign datasets (CIC-IDS-2017, CIC-IoT-2023) are public research datasets. The detector itself is a defensive system, and its published detection lower bounds (the T6 floors) let defenders assess deployment boundaries; an attacker constructing evasions along those boundaries gains no more than from directly adjusting the attack rate (Section~\ref{sec:lowerbounds}). The artifact is available at \url{https://anonymous.4open.science/r/SweepSketch-9F47}. Generative AI was used for editorial purposes in this manuscript, and all outputs were inspected by the authors to ensure accuracy and originality.

\end{document}